\documentclass[a4paper,11pt]{scrartcl}
\usepackage[a4paper,left=2.54cm,right=2.54cm,top=2.54cm,bottom=2.54cm]{geometry}
\usepackage[english]{babel}
 \usepackage[utf8]{inputenc}
\usepackage[inline]{enumitem}
\usepackage{authblk}
\usepackage[dvipsnames]{xcolor}
\usepackage{hyperref}
\usepackage{amsmath,amssymb,amsthm}
\usepackage[]{latexsym}
\usepackage[]{graphicx}
\usepackage{bbm}
\usepackage{subfigure} 
\usepackage{epstopdf}
\usepackage{csquotes}
\usepackage{lmodern}
\usepackage{tabularx}

\usepackage{tkz-berge}
\usepackage{caption}

\usepackage[titletoc,title]{appendix}

\tikzstyle{arrow}=[draw, -latex]
\usetikzlibrary{decorations.pathmorphing}

\newcommand{\var}{\operatorname{VaR}}
\newcommand{\avar}{\operatorname{AVaR}}

\newcommand{\bbr}{\mathbb{R}}
\newcommand{\bbe}{\mathbb{E}}
\newcommand{\bbn}{\mathbb{N}}

\newcommand{\fil}{\mathcal{F}}
\newcommand{\fcal}{\mathcal{F}}

\newcommand{\rcal}{\mathcal{R}}
\newcommand{\scal}{\mathcal{S}}
\newcommand{\ncal}{\mathcal{N}}

\newcommand{\ccal}{\mathcal{C}}

\newcommand{\acal}{\mathcal{A}}

\newcommand{\ycal}{\mathcal{Y}}

\newtheorem{lemma}{Lemma}[section]

\newtheorem{theorem}[lemma]{Theorem}

\newtheorem{definition}[lemma]{Definition}
\newtheorem{example}[lemma]{Example}

\newtheorem{algorithm}[lemma]{Algorithm}

\newcommand{\new}[1]{\textcolor{red}{#1}}

\usepackage{framed, xcolor}
 \definecolor{shadecolor}{gray}{0.9}

\usepackage[style=bwl-FU, doi=false,isbn=false, url=false,dashed=false, giveninits=true,uniquename=init,maxbibnames=99, backend=bibtex]{biblatex}
\usepackage{multirow,array}
\usepackage{graphicx}
\usepackage[printonlyused,nohyperlinks]{acronym}
\newcommand\MyBox[1]{%
    \fbox{\parbox[c][2cm][c]{5cm}{\centering #1}}%
}
\newcommand\MyVBox[1]{%
    \parbox[c][1cm][c]{1cm}{\centering\bfseries #1}%
}  
\newcommand\MyHBox[2][\dimexpr 5cm+2\fboxsep\relax]{%
    \parbox[c][1cm][c]{#1}{\centering\bfseries #2}%
}  
\newcommand\MyTBox[4]{%
    \MyVBox{#1}
    \MyBox{#2}\hspace*{-\fboxrule}%
    \MyBox{#3}\par\vspace{-\fboxrule}%
}  

\begin{document}
	
	\title{Modeling and Pricing Cyber Insurance\\ --\\ Idiosyncratic, Systematic, and Systemic Risks}

	\author[a,b]{Kerstin Awiszus}
	\author[a,c]{Thomas Knispel}
	\author[a,d]{Irina Penner}
	\author[a,e]{Gregor Svindland}
	\author[a,e]{Alexander Vo\ss}
	\author[a,e]{Stefan Weber}
	\affil[a]{House of Insurance, Leibniz Universit\"at Hannover}
	\affil[b]{Hochschule Hannover}
	\affil[c]{Berlin School of Economics and Law}
	\affil[d]{HTW, University of Applied Sciences, Berlin}
	\affil[e]{Institute of Actuarial and Financial Mathematics, Leibniz Universit\"at Hannover}

	\date{\today}
	
	\maketitle
	
	\begin{abstract}
		The paper provides a comprehensive overview of modeling and pricing cyber insurance and includes clear and easily understandable explanations of the underlying mathematical concepts. We distinguish  three main types of cyber risks: idiosyncratic, systematic, and systemic cyber risks. While for idiosyncratic and systematic cyber risks, classical actuarial and financial mathematics appear to be well-suited, systemic cyber risks require more sophisticated approaches that capture both network and strategic interactions. In the context of pricing cyber insurance policies, issues of interdependence arise for both systematic and systemic cyber risks; classical actuarial valuation needs to be extended to include more complex methods, such as concepts of risk-neutral valuation and (set-valued) monetary risk measures.
	\end{abstract}\vspace{0.2cm}
	\textsf{\textbf{Keywords:}}\footnote{We are grateful for useful comments of the reviewers and the editor.}  Cyber Risks; Cyber Insurance; Idiosyncratic Risk; Systematic Risk; Systemic Risk.

	
	\section{Introduction}\label{sec:intro}
	
	Cyber risks constitute a major threat to companies worldwide.\footnote{For example, according to the annually published Allianz Risk Barometer (see, e.g., \cite{allianz2022}), cyber risk ranges among the top three global business risks since 2016.} In the last years, the estimated costs of cyber crime have continuously been increasing -- from approximately USD 600 billion in 2018 to more than USD 1 trillion in 2020, cf. \cite{csis2020}. 
	Consequently, the market for cyber insurance is experiencing strong growth, providing contracts that mitigate the increasing risk exposure -- with significant potential ahead. However, cyber insurance differs from other lines of business in multiple ways that pose significant challenges to insurance companies offering cyber coverage:
	\begin{itemize}
	    \item \emph{Data} on cyber events and losses is scarce and typically not available in the desired amount or granularity.
	    \item Cyber threats are evolving dynamically in a highly \emph{non-stationary} cyber risk landscape.
	    \item \emph{Aggregate cyber risks} arise due to common IT architectures or complex interconnections that cannot easily be captured.
	    \item The term `cyber' risk itself comprises many \emph{different types of risk} with different root causes and types of impact.
	\end{itemize}   
	Insurance companies cannot solely rely on standard actuarial approaches when modeling and pricing cyber risks. Their traditional methods need to be complemented by novel and innovative techniques for both underwriting and quantitative risk management. The current paper provides the following main contributions:
    \begin{enumerate}
        \item[(i)] We present a \emph{comprehensive overview of the state of the art of modeling and pricing cyber insurance}. In contrast to other surveys (see, e.g., \cite{eling2020cyber}) that focus on a high-level review of the literature, we explain the underlying mathematical concepts and discuss their advantages and drawbacks.\footnote{Surveys that include detailed conceptual explanations are, e.g., \cite{Boehme2010}, \cite{Marotta2017}, and \cite{Boehme2018}. In contrast to our paper, these authors focus exclusively on game-theoretic models. We discuss this dimension in Section \ref{subsec:strategicinteraction}.}         
        \item[(ii)] The second main contribution of the paper is a classification of cyber risks into three different types: \emph{idiosyncratic, systematic, and systemic cyber risks}. While the distinction between idiosyncratic and systemic risks is common in the current cyber insurance literature (see, e.g., \cite{zeller2020comprehensive}), a further refinement is necessary. The three risk types can be described as follows:
        \begin{itemize}
    \item {\textbf{Idiosyncratic risks}} refer to cyber risks at the level of individual policyholders that are independent from risks of others parties. This might, for example, be caused by internal errors within the company. Prototypical idiosyncratic risks are independent risks in large insurance pools that allow to apply classical actuarial techniques.
	\item {\textbf{Systematic risks}} are cyber risks that result from common vulnerabilities of entities affecting different firms at the same time, e.g., firms belonging to the same industry sector or region, or firms that utilize the same software, server, or computer system. These risks can be modeled via common risk factors. In classical actuarial and financial mathematics, systematic risks include financial market risks as well as stochastic fluctuations and evolutions of mortality rates within a population. 
	\item {\textbf{Systemic risks}} are cyber risks caused by local or global contagion effects in interconnected systems or by strategic interaction. Examples are worm-type malware or supplier attacks. These risks are similar to important feedback mechanisms observed in financial crises, e.g., contagion in networks of counterparties or fire sales of stressed market participants in illiquid markets. Models include random processes with feedback, or locally and globally interacting processes. We will also include strategic interactions in this category which are studied in game theory.
	\end{itemize}
        
        Idiosyncratic and systematic cyber risks can be captured by classical approaches of actuarial and financial mathematics; systemic cyber risks require different methodologies such as epidemic network models which focus on the interconnectedness of the entities. We suggest pricing techniques that adequately incorporate interdependence for both systematic and systemic cyber risks by combining the concepts of risk-neutral valuation and risk measures.
    \end{enumerate}
	
	The paper is structured as follows. Section \ref{sec:ClassicalActuarialApproaches} reviews classical actuarial approaches. We begin with an introduction to the frequency-severity approach in the context of cyber risk and discuss how to model both idiosyncratic and systematic risks in this framework. We explain how dependence is captured in such models. Systemic cyber risks are  considered in Section \ref{sec:systemic}. Three different modeling approaches for interconnectedness, contagion, and interaction between entities are discussed, with a special focus on their advantages and possible drawbacks.
	In Section \ref{sec:Pricing}, we describe pricing methods for cyber insurance contracts that are applicable in the face of idiosyncratic, systematic, and systemic risks. Section \ref{sec:Conclusion} discusses open questions for future research.

	
\section{Classical Actuarial Approaches Applied to Cyber Risks}
\label{sec:ClassicalActuarialApproaches}
The pricing of cyber insurance contracts as well as quantitative cyber risk management require sound models for the loss distributions, customized to the application purpose. While classical actuarial premium principles are essentially related to the expected claims amount (plus a safety loading), quantitative risk management particularly refers to extreme losses in the tail of the distribution and their quantification in terms of risk measures such as \emph{Value at Risk} or \emph{Average Value at Risk}, see Section \ref{sec:Pricing}.

In actuarial mathematics, a standard model for insurance losses -- used across all lines of business -- is the \emph{frequency-severity approach}, also called \emph{collective risk model}. For a certain time interval $[0,t]$, $t>0$ (typically $t=1$ year), a collective of policyholders causes a random number of claims $\ncal_t$ (\emph{frequency}) with corresponding random loss sizes $\ycal_1, \ycal_2,\ldots$ (\emph{severity}) generating the total claim amount 
$$\scal_t=\sum_{j=1}^{\ncal_t} \ycal_j, \quad t>0.$$
Calculations within the frequency-severity approach typically rely on the following mathematical assumptions (see, e.g., \cite{mikosch2004non}):
\begin{enumerate}
\item[(C1)] Claims occur at arrival times $0\le T_1\leq T_2\leq \ldots$. The  number of claims in the time interval $[0,t]$, $t\ge 0$, is defined by
$$\ncal_t:=\#\{j\ge 1 \;\vert\; T_j\le t \}, $$
i.e., $\ncal=(\ncal_t)_{t\ge 0}$ constitutes a counting process on $[0,\infty)$.
\item[(C2)] The $j$th claim arriving at time $T_j$ causes the claim size $\ycal_j$. It is assumed that the sequence $(\ycal_j)_{j\ge 1}$ of claim sizes consists of independent and identically distributed random variables.
\item[(C3)] Claim sizes and claim numbers are assumed to be independent from each other.
\end{enumerate}
In contrast to classical insurance risks, however, cyber risk is more challenging in different ways. In particular, the standard assumptions of the frequency-severity approach as well as classical statistical techniques\footnote{For details on statistical techniques in classical actuarial models, see Sections \ref{subsec:CalEvalFreqSev} and \ref{subsec:CalEvalDepModels}.} are no longer applicable:
\begin{itemize}
\item Claims \emph{data} are not available in sufficient quantity or in the required granularity.
\item Technology and cyber threats are evolving rapidly, i.e., the cyber environment is highly \textit{non-stationary}.
\item Cyber incidents\footnote{According to \cite{NISTglossary}, a cyber incident can be defined as: \lq\lq{}Actions taken through the use of an information system or network that result in an actual or potentially adverse effect on an information system, network, and/or the information residing therein.\rq\rq{} We will also use the term cyber attacks interchangeably in this paper.}  may affect different policyholders at the same time, i.\,e., the typical assumption of \emph{independence} for insurance risks does not hold any longer. Moreover, there is -- in contrast to natural catastrophe risk -- no simple geographical delimitation of dependent risks.
\end{itemize}
Nonetheless, the frequency-severity approach can be customized to account for cyber risk -- at least as a first approximation and for certain types of non-systemic cyber risks, which can be subdivided into idiosyncratic and systematic risks (as defined in Section \ref{sec:intro}).
In the frequency-severity approaches presented below, we explicitly distinguish between techniques suitable for modeling idiosyncratic or systematic incidents. In the context of cyber insurance, however, a third class of risks can be identified, namely systemic risks, i.e., cyber risks resulting from contagion between interconnected entities.
Proper modeling of such risks goes beyond the classical framework of actuarial modeling and requires appropriate models for networks, (cyber) disease spread, and strategic interaction. Hence, we discuss the modeling of systemic cyber risks separately in Section \ref{sec:systemic}, while the pricing for all types of cyber risks is discussed in Section \ref{sec:Pricing}.
	
To present frequency-severity approaches in the context of cyber risk in a unified and practically applicable way, we use the following notation and definitions. We consider an insurer's portfolio of $n$ policyholders (firms) exposed to the considered type of cyber risk incidents. Each firm admits an individual risk profile characterized by a vector of covariates, e.g., \emph{industry sector}, \emph{size}, \emph{IT security level}, which are elicitable, for example, via a questionnaire or from public information. Using the covariates, the insurer's portfolio is decomposed into homogeneous groups, labeled $\{1,\ldots, K\}$, with covariates vector $x^k$ for group $k$.  We denote by $n_k$, $k=1,\ldots,K$, the number of firms in group $k$, i.e., $n_1+\ldots+n_K=n$. For pricing purposes, these homogeneous groups can be viewed as tariff cells, i.e., the insurance firm should charge all firms\footnote{For simplicity, we assume that the firms within a group possess the same types of exposures that are of the same size. This can be generalized by introducing suitable volume measures that characterize the size of exposures.} within group $k$ the same premium $\pi_k$. In particular, if $n_k$ is large, then the premium of the idiosyncratic cyber risk can be derived from the law of large numbers as the expected claims amount per firm of group $k$ plus a suitable safety loading to avoid ruin in the long run.

Both idiosyncratic and systematic incidents can be grouped into different cyber risk categories, labeled $\{1,\ldots, C\}$. Categories may include, for example, \emph{data breach}, \emph{fraud}, and \emph{business interruption}. Two exemplary actuarial classification approaches are sketched and discussed in Appendix \ref{sec:riskcat}.
Cyber risk is modeled per risk category $c\in\{1,\ldots,C\}$ and per group $k\in\{1,\ldots, K\}$. A pair $m:=(c,k)$ is called a \emph{cyber risk module}. The total number of modules $C\cdot K$ is a trade-off between homogeneity and availability of data for statistical estimation.

Within this framework, we model the losses for an insurance company -- for each cyber risk module as well as on an aggregate level. For this purpose, we first focus on frequency-severity based approaches to modeling cyber risks in the spirit of the classical collective risk model. Second, we add dependence to our cyber risk model in order to capture accumulation risks. Note that appropriate dependence modeling is particularly important for calculating capital requirements in quantitative risk management, since the underlying risk measures refer to events in the extreme tail of the loss distribution.

\subsection{Frequency and Severity}\label{sec:frequencyseverity}
A frequency-severity model may be applied on the level of each cyber risk module $m=(c,k)$. For simplicity, we describe the losses per risk category of individual firms by a collective risk model. This can be justified as follows: Since all firms in any group are (approximately) homogeneous, they will be charged the same premium for any given risk category. From the point of view of the insurance company, only aggregate losses are relevant, i.e., an artificial allocation of losses to individual companies for pricing purposes will produce the correct implications.  We thus describe the losses per risk category \emph{at the level of any individual firm} by a collective risk model with the same severity as the corresponding module, but with a suitably reduced frequency. 

For a firm $i$ in group $k$ and a fixed risk category $c$, i.e., a cyber risk module $m=(c,k)$, we consider the frequency and severity model $(\ncal_{t}^{m,i},(\ycal^{m,i}_j)_{j\geq 1})$. Then the total claim amount of firm $i$ up to time $t$ can easily be obtained by summing up:
$$\scal_t^{m,i}=\sum_{j=1}^{\ncal_t^{m,i} }\ycal_j^{m,i}.$$
In mathematical terms, all quantities correspond  to random variables on a suitable probability space $(\Omega,\mathcal{F},\mathbb{P})$, where $\mathbb{P}$ plays the role of the statistical measure that models the relative frequency with which events occur.

As outlined in the introduction of this section, one of the most common assumptions in the frequency-severity model is assumption  (C3), i.\,e., claim numbers and sizes are independent of each other. This assumption facilitates and simplifies many calculations regarding the compound total claim amount process. In particular, the expected total claim amount and its variance follow from \emph{Wald's formulas:}
$$\bbe[\scal_t^{m,i}]=\bbe[\ncal_t^{m,i}]\cdot\bbe[\ycal_1^{m,i}], \quad \operatorname{Var}(\scal_t^{m,i})=\bbe[\ncal_t^{m,i}]\operatorname{Var}(\ycal_1^{m,i})+\operatorname{Var}(\ncal_t^{m,i})(\bbe[\ycal^{m,i}_1])^2.$$
However, the independence assumption may not always be reasonable  -- e.g., if hidden factors influence both frequency and severity: \cite{Sun2020} detect a positive nonlinear dependence between frequency and severity in hacking breach risks on firm-level. A firm with a strong cyber self protection is expected to experience both fewer and weaker hacking attacks than companies with weak self protection mechanisms. In mathematical terms, the authors capture this dependence between frequency and severity by the Gumbel copula, see also Section \ref{subsec:dependence}.

\subsubsection{Frequency}\label{subsec:frequency}
Let $\ncal_t^{m,i}$ denote the number of incidents in module $m=(c,k)$ until time $t$ that are allocated to a firm $i$ in group $k$, and let $(\ncal_t^{m,i})_{t\geq 0}$ denote the corresponding counting process. At the aggregate level, $$\ncal_t^{m, agg}:=\sum_{i=1}^{n_k}\mathcal{N}_t^{m,i}\mbox{ and } \ncal_t^{(c)}:=\sum_{k=1}^K\mathcal{N}_t^{m,agg},\quad t\geq 0,$$
will count the total number of incidents per module $m= (c,k)$ and the total number of incidents per cyber risk category $c$, respectively.

\paragraph{Poisson Process} A simple counting process for incidents -- reflecting non-stationarity of cyber risk -- is a \emph{time-inhomogeneous} Poisson process with intensity function $\lambda^m$ per firm for cyber risk module $m$. 
\begin{definition}[Time-inhomogeneous Poisson process] A counting process $(\ncal_t)_{t \geq 0}$ is called a time-inhomogeneous Poisson process on $(\Omega,\mathcal{F},\mathbb{P})$ with locally integrable rate (or intensity) function $\lambda:[0,\infty)\rightarrow [0,\infty)$ if:
\begin{enumerate}
\item $\ncal_0= 0$,
\item the process has independent increments,
\item for any time interval $(s,t]$, the number of incidents is Poisson distributed with mean $\int_s^t\lambda(u)\,du$, i.e.,
$$\ncal_t-\ncal_s\sim \operatorname{Poiss}\left(\int_s^t\lambda(u)\,du\right).$$
\end{enumerate}
\end{definition}
Unless the intensity function is constant, the increments of a time-inhomogeneous Poisson process are \emph{non-stationary}. The cumulative rate function $\int_0^t\lambda(u)\,du$ corresponds to the expected number of incidents up to time $t$.  

\cite{zeller2020comprehensive} adopt this approach for idiosyncratic incidents. For each policyholder $i$ of group $k$ and module $m=(c,k)$, the number of idiosyncratic incidents $(\ncal_t^{m,i})_{t\geq 0}$ is assumed to follow a time-inhomogeneous Poisson process with intensity $\lambda^m=\lambda^{(c,k)}$. Clearly, for each cyber risk category $c$, the intensity at the level of an individual firm $i$ depends on the covariates $x^k$ of group $k$ (but not on the individual policyholder $i$), and \cite{zeller2020comprehensive}  propose a \emph{generalized additive model}
$$\lambda^{(c,k)}(t)=\exp(f^c(x^k)+g^c(t))$$
to estimate the intensity rates.\footnote{The auxiliary function $f$ additively maps the covariates, while  $g$ captures the time dependence.} In particular, similarities and deviations of the risk profiles of the $K$ groups -- expressed in terms of the covariate vectors $x^k$, $k=1,\ldots,K$ -- are reflected by the intensity functions $\lambda^{(c,k)}$.
 
 Since \emph{idiosyncratic} incidents are independent across firms, the total number of incidents $\ncal_t^{m, agg}$, $t\geq 0$, per module $m= (c,k)$ as well as the total number of incidents $\ncal_t^{(c)}$, $t\geq 0$, per cyber risk category $c$, respectively, are again  time-inhomogeneous Poisson processes with respective intensities
\begin{equation}
\label{eq:PoissonAdditivity}
\lambda^{m, agg}(t)=  n_k \lambda^{(c,k)}(t),\quad\lambda^{(c)}(t)=\sum_{k=1}^K n_k \lambda^{(c,k)}(t),     \quad t\geq 0.
\end{equation}
More delicate, however, is the case of \emph{systematic} cyber risk incidents. In particular, frequency distributions of different policyholders might be subject to dependencies due to joint underlying cyber risk factors $\rcal^1,\ldots,\rcal^d$,  representing, for example, the random discovery of exploits in commonly used software,  improvements in cyber security, or the technological progress of tools for cyber attacks.

\paragraph{Cox Process}  Such dependencies between counting processes can be captured in the context of \emph{Cox processes}, also called \emph{doubly stochastic Poisson processes}, extending the notion of a time-inhomogeneous Poisson process to a random intensity. 
\begin{definition}[Cox process]
A Cox process $(\ncal_t)_{t\geq 0}$ is a counting process described by a random intensity process $(\lambda_t)_{t\geq 0}$ such that conditional on the specific realization $t\mapsto \lambda_t(\omega)$, $\omega\in\Omega$, the process $(\ncal_t)_{t\geq 0}$ is a time-inhomogeneous Poisson process with intensity  $t\mapsto \lambda(t)=\lambda_t(\omega)$. 
\end{definition}

A reasonable assumption could be that the intensity is a function of the current state of random cyber risk factors, i.e., for an $\mathbb{R}^d$-valued stochastic process $\rcal_t=(\rcal^1_t,\ldots,\rcal_t^d)$, $t\geq 0$, of cyber risk factors and a function $\lambda: \mathbb{R}^d\rightarrow [0,\infty)$, the intensity process is defined as
$$\lambda_t(\omega)=\lambda(\rcal_t(\omega)), \quad t\geq 0,\,\omega \in\Omega.$$
More generally, the intensity process could be modeled as a function of the whole history of cyber risk factors, i.e., 
$$\lambda_t(\omega)=\lambda(\rcal_u(\omega):u\leq t), \quad t\geq 0,\,\omega \in\Omega.$$
 In summary, in the case of systematic cyber risk, a reasonable model for the number of incidents $\ncal_t^{m,i}$ up to time $t$ allocated to policyholder $i$ in group $k$ for module $m=(c,k)$ could be to assume that $(\ncal_t^{m,i})_{t\geq 0}$  follows a Cox process with intensity process $\lambda_t^m=\lambda^m(\rcal_t)$, $t\geq 0$, defined in terms of a suitable function $\lambda^m:\mathbb{R}^d\rightarrow \infty$, such that conditional on the cyber risk factors $t\mapsto \rcal_t(\omega)=(\rcal^1_t(\omega),\ldots, \rcal_t^d(\omega))$ the counting processes $(\ncal_t^{m,i})_{t\geq 0}$, $m = (c,k)$, $c =1,\ldots,C$, $k=1, \ldots, K$, are independent time-inhomogeneous Poisson processes. In particular, conditional independence implies that -- conditional on the specific realization $t\mapsto \lambda^m_t(\omega)$ -- the total number of incidents $\ncal_t^{m, agg}$, $t\geq 0$, per module $m= (c,k)$ and the total number of incidents $\ncal_t^{(c)}$, $t\geq 0$, per cyber risk category $c$ are again time-inhomogeneous Poisson processes with intensities
 $$\lambda^{m, agg}_t=  n_k \lambda^{(c,k)}_t,\quad\lambda^{(c)}_t=\sum_{k=1}^K n_k \lambda^{(c,k)}_t,     \quad t\geq 0,$$
in analogy to (\ref{eq:PoissonAdditivity}).

In contrast to the time-inhomogeneous Poisson process, the increments of a Cox process $(\ncal_t)_{t\geq 0}$ are in general no longer independent, but subject to autocorrelation. More precisely, for any $s<t\leq u<v$, the tower property of conditional expectation implies
$$ \operatorname{Cov}(\ncal_t-\ncal_s,\ncal_v-\ncal_u)=\operatorname{Cov}\left(\int_s^t\lambda_z\,dz,\int_u^v\lambda_z\,dz\right),$$
i.e., the autocorrelation depends on the random intensity process. The statistical analysis of \cite{bessy2020multivariate} yields empirical evidence for autocorrelation in the number of attacks, and thus provides an additional rationale for Cox processes when modeling claims frequency. The specification of an intensity process that reproduces the empirically observed autocorrelation appears to be challenging.


\subsubsection{Severity}\label{subsec:classicalseverity}
Every claim occurring in the frequency-severity model triggers a loss size that is modeled as a random variable. We let $\ycal_{j}^{m,i}$ denote the claim size of the $j$th event allocated to firm $i$ for module $m=(c,k)$ and assume that $(\ycal_j^{m,i})_{j\geq 1}$, $i=1, \dots, n_k$, is a collection of non-negative independent\footnote{Cyber event claim sizes in a certain time interval may not always be independent, e.g., due to commonly used cyber security measures. The resulting dependence structures could be captured by alternatively imposing \emph{conditional} independence assumptions given a set of joint underlying risk factors -- similar to the conceptual idea underlying Cox processes that we already discussed above.} and identically distributed random variables. One among many different possible approaches is to assume that the key governing parameter for the choice of the claim size distribution is the incident category $c$; characteristics of group $k$ then determine distributional details, e.g., parameter values. 

Due to the limited availability of loss data, empirical research on cyber risk severity distributions has mostly focused on the category of data breaches. For this category, open source data bases, such as the Privacy Rights Clearinghouse Chronology of Data Breaches, are available and regularly updated. Data breach severities are found to follow strongly heavy-tailed distributions such as power-law (see, e.g., \cite{maillart2010heavy}), log-normal (see, e.g., \cite{edwards2016hype}) or generalized Pareto distributions (GPD) (see, e.g., \cite{wheatley2016extreme} or \cite{Sun2020}). For cyber risk categories different from data breaches, less data is publicly available. Consequently, fewer papers have appeared that empirically analyze the respective severity distributions. 

An exception is \cite{DaKr2020} who study a non-public database of the \emph{French Gendarmerie Nationale} on cyber complaints and describe a process for cleaning the data. Their analysis suggests that losses are heavy-tailed. \cite{DDK2022} refine the analysis and provides a tool for classifying attacks based on the fatness of the tail. Another promising direction are studies based on data on operational  risk such as \cite{biener2015insurability} or \cite{eling2019actual}. These approaches offer the benefit of being able to analyze all categories of cyber incidents simultaneously.  In particular, \cite{eling2019actual} detect distributional differences between small and large claim sizes for all considered cyber incident categories. The authors propose a \emph{composite distribution approach}, where excess losses over a threshold are modeled using a GPD and the remaining smaller losses are modeled using a simple parametric distribution such as a gamma or log-normal distribution. In general, {composite distribution approaches} constitute a flexible modeling tool to take the empirically observed distributional differences between body and tail of severity distributions adequately into account. A {composite distribution approach} can be formalized as follows. 

For each module $m$, we choose a threshold $\theta^m$ distinguishing small from large cyber claims. Small and large claims, i.e., the body and tail of the severity distribution, are then modeled separately: The i.i.d. claim sizes follow a {composite distribution} with density
\[f_{\ycal_j^m}(y):=\begin{cases}
	{C_1^m}\cdot f_{\text{small}}^m(y),&\quad\text{if $-\infty<y\le \theta^m$},\\
	{C_2^m}\cdot f_{\text{large}}^m(y),&\quad\text{if $\theta^m<y<\infty$},
\end{cases} \]
where $f_{\text{small}}^m, f_{\text{large}}^m$ are probability density functions modeling the sizes of small and large claims in module $m$, respectively, and $C_1^m$, $C_2^m$ are normalizing constants that are additionally constrained by continuity conditions at the threshold $\theta^m$.
Depending on the characteristics of the module $m$, different choices for $f_{\text{small}}^m, f_{\text{large}}^m$ may be suitable. Examples include
\begin{itemize}
\item \textbf{Small Claims:} PERT, Normal, Gamma, Log-Normal, GPD, Kernel Distribution
\item \textbf{Large Claims:} GPD
\end{itemize}

The composite distribution approach is well-suited for modeling non-life insurance severity distributions in general, and cyber risks in particular.\footnote{\cite{Sun2020} also suggest a composite distribution approach for modeling malicious hacking data breach risk. The tail of their distributions follows a GPD, and the distribution body is modeled using a non-parametric kernel distribution. Due to both its suitability and flexibility, a similar approach is also incorporated in the cyber risk model of \cite{zeller2020comprehensive}.} As discussed here, the methodology is independent of time, i.e., it provides only a snapshot of the current cyber environment. In the light of the fast-evolving, non-stationary cyber landscape, the suitability of the model must, however, be regularly validated and updated. For further details and discussions, we refer the interested reader to the excellent summaries provided by \cite{zeller2020comprehensive}, Section 2.1, or \cite{eling2020cyber}, in particular Tables 4 and 6, and to \cite{coorayComposite} for an application of composite distributions in a non-cyber specific context.

\subsubsection{On Calibration and Application}\label{subsec:CalEvalFreqSev}
In general, frequency-severity models are well-understood, easy to implement and to calibrate if a sufficient amount of data is available. They are also straightforward to explain, for example, to an executive board of an insurance company; this is partly due to their prevalence in actuarial modeling. For frequency modeling, intensities can, e.g., be fit to data using generalized additive models (as in \cite{zeller2020comprehensive} and described above), maximum- or marginal likelihood, or Bayesian methods. Cox processes are generally more difficult to estimate -- the choice of a calibration method critically depends on the law of the underlying common risk factor processes.\footnote{For details on the statistical estimation of point processes and theoretical background see, e.g., \cite{daley2003introduction}.}

For the statistical analysis of the severity, there exist well-known estimation techniques including maximum-likelihood, see, e.g., \cite{maillart2010heavy} or \cite{edwards2016hype} for applications in a cyber severity context, or the peaks-over-threshold method for fitting a GPD to the tail of a distribution, see, e.g., \cite{mcneil2015quantitative} and \cite{embrechts2013modelling}. For a general review on methods for the parameter estimation of GPDs, including maximum-likelihood, the method of moments, the probability weighted moments method, and Bayesian approaches, see also \cite{Bermudez2010} and \cite{Bermudez20102}.

The practical application of frequency-severity models to cyber risk is challenging, in particular due to the limited amount of available data and their insufficient quality. Moreover, Poisson and Cox processes do not capture the systemic interaction between different (groups of) policyholders; see also \cite{reinhart2022discussion} for a discussion of the frequency-severity model presented by \cite{zeller2020comprehensive}. An alternative are Hawkes processes that incorporate systemic self-excitation into frequency models, see Section \ref{subsec:Hawkes}. Like Cox processes, Hawkes processes are able to capture autocorrelation observable in the data.

\subsection{Dependence Modeling}\label{subsec:dependence}
The distribution of the total claim amount per module and at the portfolio level is affected by the underlying dependence structures. For cyber risk, dependencies may be present at different levels including:
\begin{itemize}
\item dependence between frequency distributions or between severity distributions of different policyholders in the same homogeneous group (e.g., due to the random evolution of common cyber security measures and cyber threats over time),
\item dependence between frequency and severity -- in contrast to the classical framework of frequency-severity models (e.g., due to unobservable random factors within a tariff class such as heterogeneous levels of cyber self protection).
\end{itemize}
One approach to deal with the first type of dependencies are Cox processes as described in Section \ref{subsec:frequency}. In this section, we review further approaches to model dependence in the context of cyber risk that have been proposed in the literature. 

\subsubsection{Common Risk Factors}\label{subsubsec:commonfactor}

Common risk factors capture dependence for systematic risks; the factors are random quantities to which all risks are jointly exposed. Common risk factors appear in static as well as in dynamic models and have been widely used in the cyber risk modeling literature. For example, they are key elements of the cyber risk models proposed by \cite{bohme2005cyber}, \cite{bohme2006models} and \cite{zeller2020comprehensive}. Cox processes, as introduced in Section \ref{subsec:frequency}, are an example of dynamic factor models.

\cite{bohme2005cyber} captures dependence using one common risk factor in a static model. The factor represents a common vulnerability in a portfolio of $n$ individual risks. The connection between individual risks and the latent risk factor is studied on the basis of linear correlation.\footnote{Linear correlation, defined as $\rho(X,Y)={\textrm{Cov}(X,Y)}/{\sqrt{\textrm{Var}(X)\textrm{Var}(Y)}}\in[-1,1]$, captures a possible linear relationship between the random variables $X$ and $Y$. The maximum and minimum values of $1$ and $-1$ are not always attainable. While often used to impose ad-hoc dependence assumptions in practice (in a cyber context, see, e.g., \cite{bohme2005cyber}, \cite{bohme2006models}), linear correlation suffers from many well-known fallacies, see, e.g., \cite{mcneil2015quantitative} for a detailed discussion.} Common risk factors also appear in the cyber risk model of \cite{zeller2020comprehensive}. The authors use marked point processes with two-dimensional marks: the first component describes the strength of an attack, and the second component represents the subset of companies affected. Dependence among firms occurs due to the restriction of incidents to certain industry sectors which is modeled via a common risk factor. The paper suggests a conceptual framework, but does not yet calibrate the model to real data.

\subsubsection{Copulas}\label{subsec:copulas}
In actuarial applications, copulas are a standard tool that fully characterizes the dependence structure of the components of finite-dimensional random vectors. A $d$-dimensional \emph{copula} $\ccal:[0,1]^d\rightarrow[0,1]$ is the distribution function of a $d$-dimensional random vector with uniform one-dimensional marginal distributions. 
\begin{theorem}[Sklar's Theorem]
\begin{itemize}
\item[1.]
For any $d$-dimensional distribution function $F$ with margins $F_1,\ldots,F_d$ there exists a copula $\ccal$ with
\begin{equation} 
\label{eq:SklarTheorem}
F(x_1,\ldots,x_d)=\ccal(F_1(x_1),\ldots,F_d(x_d))\quad \mbox{for all $x_1,\ldots,x_d\in[-\infty,\infty]$.}
\end{equation}
If all $F_i$ are continuous, then $\ccal$ is unique. 
\item[2.] Conversely, for a given copula~~$\ccal$~and given one-dimensional distribution functions $F_1,\ldots,F_d$, the function $F$ in 
(\ref{eq:SklarTheorem}) is a $d$-dimensional distribution function with copula $\ccal$ and marginal distribution functions $F_1,\ldots,F_d$.
\end{itemize}
\end{theorem}
Property 1 states that a copula extracts the dependence structure of a random vector from its multivariate distribution, while property 2 provides a flexible construction principle of multivariate models by combing marginal distributions and copulas to multivariate distributions. Prominent examples of copulas are:
\begin{itemize}
\item \textbf{Gaussian copula:} Letting $\Phi^{-1}$ be the quantile function of the standard normal distribution and $\Phi_{\Sigma}$ the joint cumulative distribution function of a multivariate normal distribution with covariance matrix $\Sigma$, the corresponding Gaussian copula is given by
\begin{equation*}
\ccal_\Sigma^{\mathrm{Ga}}(u_1,\ldots , u_{d}) = \Phi_{\Sigma} (\Phi^{-1}(u_1),\ldots , \Phi^{-1}(u_{d})) \quad ({(u_1,\ldots,u_d)}\in[0,1]).
\end{equation*}
\item \textbf{$t$-copula:} Let $t_{\nu,\Sigma}$ signify the distribution function of a $d$-dimensional $t$-distribution $t_d(\nu,0,\Sigma )$ for a given correlation matrix $\Sigma$ and with $\nu$ degrees of freedom, and let $t_\nu$ denote the distribution function of a univariate $t$-distribution with $\nu$ degrees of freedom. The corresponding $t$-copula takes the form
$$\ccal^t_{\nu,\Sigma}(u_1,\ldots,u_d)=t_{\nu,\Sigma}(t_\nu^{-1}((u_1),\ldots,t_\nu^{-1}(u_{d}))\quad ({(u_1,\ldots,u_d)}\in[0,1]).$$
Like the Gaussian copula, the $t$-copula is an implicit copula that is extracted from a given parametric multi-variate distribution.
\item \textbf{Archimedean copulas}: Explicit copulas are constructed from given functions; the prime example are Archimedean copulas. We consider a suitable continuous function $\psi:[0,\infty )\to [0,1]$ with $\psi(0) = 1$, $\lim_{x\to\infty} \psi(x) = 0$, and $\psi$ strictly decreasing on $[0, \psi^{-1}(0)]$, where $\psi^{-1}$ denotes its generalized inverse. The Archimedean copula with generator $\psi$ is given by
\begin{equation*}
\ccal_\psi^{\mathrm{Ar}}(u_1,\ldots , u_{d}) = \psi^{-1} (\psi (u_1) +\ldots + \psi(u_{d})) \quad ({(u_1,\ldots,u_{d})}\in[0,1]).
\end{equation*}
A special case is the \textit{Gumbel copula} for $\psi_\theta (s) = (-\ln (s))^\theta$, $\theta \in [1,\infty )$ that is applied in the cyber model of \cite{Sun2020}.
 \end{itemize}

\subsubsection{On Calibration and Application}\label{subsec:CalEvalDepModels}

Common risk factor models are able to capture dependence from bottom-up and are widely used in economics. From a practical perspective, they are particularly useful when a modeler is confident that random outcomes are influenced by common external factors. In Cox processes, described in Section \ref{subsec:frequency},  the common factors enter the model via the intensity. Their estimation  depends on the specific choice of the distribution of the underlying risk factors. For the class of \emph{linear} factor models, a large amount of statistical estimation methods exist. Important techniques are time series regression, cross-sectional regression (at each time point), and principal component analysis, see, e.g., \cite{mcneil2015quantitative} and the references therein.

Another approach are copulas; these are theoretically able to represent every form of static dependence. They can be viewed as a top-down approach that imposes a dependence structure without modeling the underlying mechanisms, as contrasted with factor models that can be interpreted as a bottom approach. Copulas have already been used in the literature on cyber risk.   \cite{herath2011copula} model the loss distribution \emph{at a single firm} using a copula that captures the dependence structure between the number of affected computers of the firm and the overall severity of the loss. In \cite{bohme2006models} dependence \emph{between different firms} is captured using a $t$-copula with a given linear correlation coefficient. 

Another example is an application of copulas in a modified collective risk model in which the standard independence assumption is relaxed.  For the incident category $c$ of {hacking data breaches}, \cite{Sun2020} observe upper tail dependence between frequency and severity. This may be caused by hidden factors such as the degree of cyber self protection. They propose to model this dependence for any firm $i$ in module $m$ up to time $t$ via a Gumbel copula.

\cite{eling2018copula} and \cite{liu_li_daly_2022} apply vine copulas in the context of data breaches. Vine copulas are very flexible, and their calibration is quite tractable, since high-dimensional dependence structures are decomposed into components of lower dimension. For detailed information on vine copulas we refer to \cite{czado2019analyzing},  \cite{czadonagler2022}, and an online collection of material on vine copulas, see \cite{TUvinecopulapage}.

In general, the choice of a suitable copula estimation method depends on the structure of the chosen copula model: parametric, semiparametric or nonparametric.  A good survey on various methods is \cite{HKMY18}.
In a fully parametric model, both the copula and the marginal distributions are completely characterized by (vector) parameters. The maximum likelihood (ML) method can be applied to the dependence and the marginal part either jointly or sequentially. The sequential approach is often referred to as the method of inference functions for margins (IFM), see, e.g., the surveys in \cite{Choros2010}, Section 2.1, or \cite{mcneil2015quantitative}, Section 7.6.
Semiparametric approaches typically still involve a parametric copula model, but a nonparametric model for the marginals. Here, classically, the marginal distributions are estimated via their empirical distribution functions. Estimation of the full model can then be performed using a maximum-pseudo likelihood approach, in which the nonparametric marginal estimators are inserted, see the seminal paper of \cite{Genest1995semiparametric}. This approach is considered to be more robust than the parametric ML and IFM methods in many practical applications, see \cite{Kim2007}, unless substantial information is available on a parametric class to which the margins belong to.
Nonparametric copula models may be estimated on the basis of different variations of nonparametric marginal and joint distribution function estimates, see, e.g., the seminal paper of \cite{deheuvels1979} using empirical distribution functions or \cite{Chen2007} (and the references therein) for kernel-based estimators of the copula (or copula density).	

\section{Systemic Cyber Risks}\label{sec:systemic}
Systemic risk generally refers to the possibility that distortions in a system may spread across many entities and be augmented due to local or global feedback effects. This is in contrast to systematic risk that introduces dependence via exogenous factors. Systemic risk refers to the internal mechanism of a system in which the behavior of the various entities has a sequential impact. It is often associated with a cascading risk propagation such that 
\begin{quote}
``in case of an adverse local shock (infection) to a system of interconnected entities, a substantial part of the system, or even the whole system, finally becomes infected due to contagion effects.''\footnote{See \cite{Detering2019}.}
\end{quote}
As a consequence of the 2008 financial crisis, systemic risk was intensively studied in systems of interdependent financial institutions, see, e.g., \cite{staum2013systemic}. This concept is also important in the context of cyber risk, since agents and organizations in cyber systems are interconnected, for example within IT networks or via business contacts.\footnote{See, e.g., the discussion in Section 2 of \cite{Welburn2021}.} The relevance of systemic cyber threats has been emphasized by leading regulatory and macroprudential institutions, cf.~\cite{WEF2020} and \cite{Euro2020}. Examples of contagious threats include the WannaCry and NotPetya cyber attacks where the corresponding malware spread through networks of interconnected IT devices and firms, causing tremendous losses to cyber systems worldwide.\footnote{For further information and a detailed risk analysis see \cite{Euro2020}.}

Modeling systemic cyber risks requires models of feedback effects, local and global interaction, as well as strategic interaction. We describe three concrete methodological approaches (see Figure \ref{fig:interaction}): Firstly, self-excitation of cyber incidents can be captured by \textit{Hawkes processes} on an aggregate level (Section \ref{subsec:Hawkes}); in this respect, Hawkes processes can be interpreted as a top-down approach.  Secondly, \textit{epidemic network models} (Section \ref{subsec:interconnectednetworkmodels}) capture the interconnectedness and cascading propagation of risks;  this bottom-up approach may focus on local connections, but can also capture global interaction via aggregate, mean-field quantities. Both approaches can be viewed as mechanistic interaction models in which rational or strategic behavior of agents is typically not  
mirrored. This is the focus of the third approach, \textit{game-theoretic models} (Section \ref{subsec:strategicinteraction}). These study explicitly the \textit{strategic} interaction of interconnected entities, usually under strongly simplified connectivity assumptions; notions of equilibria typically characterize the solutions.

\begin{figure}[h]
    \begin{center}
    {

        \offinterlineskip

       \hspace*{0cm}\MyHBox[\dimexpr3.4cm+6\fboxsep\relax]{Interaction}\par

        {\hspace*{0cm}\MyHBox{mechanistic}\MyHBox{strategic}}\par

        \MyTBox{}{Epidemic Network Models Hawkes Processes}{Game-Theoretic Models}

    }
\end{center}
\caption{Interaction in models of systemic cyber risks}
\label{fig:interaction}
\end{figure}

	
\subsection{Hawkes Processes}\label{subsec:Hawkes}
\emph{Systematic} dependence of cyber incidents can be modeled by Cox processes; these permit to capture empirical features such as the autocorrelation of cyber attacks. Cox processes focus on common factors, but they do not model contagion in interconnected systems. An alternative are \emph{Hawkes processes}, self-exciting processes, that mirror feedback effects, a specific form of \textit{systemic} cyber risk; they also capture the stylized fact of autocorrelation of the number of events.  
\begin{definition}[Hawkes process]
A one-dimensional Hawkes process $(\ncal_t)_{t\geq 0}$ is a point process with jump times $T_1,T_2,\ldots$ and with random intensity $t\mapsto\lambda_t$, given by
$$\lambda_t=\mu(t)+\sum_{T_n\leq t}\varphi(t-T_n)=\mu(t)+\int_{[0,t)}\varphi(t-u)\,d\ncal_u,$$
where $\mu(\cdot)$ is a baseline intensity of jumps,  and where $\varphi$ is the excitation function or kernel function resp.\,which expresses the positive influence of past incidents at time $T_n$ on the current value of the intensity. 
\end{definition}
From a conceptual point of view, Hawkes processes allow to capture -- besides autocorrelation of the number of cyber risk incidents -- excitation effects, by coupling the arrival rate of events with the number of past incidents. In particular, this allows modeling systemic incidents that affect a very large number of counterparties at the same time, e.g., the spread of worm-type malware.

Self-excitation of cyber incidents for each policyholder as well as the excitation between policyholders of different groups can be modeled by a multivariate Hawkes model. More precisely, for all cyber risk modules $m=(c,k)$ and for any policyholder $i$ of group $k$, the intensity of the counting process $(\ncal_t^{m,i})_{t\geq 0}$ takes the form
 $$\lambda^{(c,k,i)}_t=\mu^{(c,k)}(t)+\sum_{l=1}^K\sum_{j=1}^{n_l}\sum_{T^{(c,l,j)}_n\leq t}\varphi^{c,k,l}_{i,j}(t-T^{(c,l,j)}_n),$$
 where
 \begin{itemize}
 \item $t\mapsto \mu^{(c,k)}(t)$ is the deterministic base intensity function, depending on the cyber risk module $m=(c,k)$ only,
 \item $t\mapsto \varphi^{c,k,l}_{i,j}(t)$ are self- and mutually-exciting maps (called kernels), depending on both the cyber risk module $m=(c,k)$, the other group $l$ and the individual policyholders $i,j$,
 \item and $T_n^{(c,l,j)}$, $n\in\mathbb{N}$, are the claims arrival times of policyholder $j$ in group $l$ with respect to the cyber risk category $c$.
 \end{itemize}
In this multivariate Hawkes model, the kernels $\varphi^{c,k,k}_{i,i}$ describe the self-excitation for policyholder $i$ of group $k$, while the $\varphi^{c,k,l}_{i,j}$ for different policyholders $i\not =j$ model contagion between policyholders and across groups.

\subsubsection{On Calibration and Application}
Using suitable parametric functions for both the baseline intensity and the kernels of Hawkes processes can in principle be estimated by maximum-likelihood methods -- provided that data is available in the desired amount and granularity. Data availability is, of course, still a major challenge in cyber insurance. Model calibration and statistical parameter estimates in a cyber context are, e.g., presented in \cite{bessy2020multivariate} focusing on data breaches. Further, Hawkes processes are also used in an empirical study of cyber risk contagion in \cite{Baldwin2017}. In the context of financial data, maximum-likelihood methods and graphical goodness-of-fit are, e.g., discussed in \cite{embrechts2011hawkes}. \cite{FoZa14} develop an estimation by the method of moments which is fast compared to likelihood estimation. A general discussion including Bayesian estimation is presented in \cite{daley2003introduction}, see also \cite{giesecke2008credit}, \cite{EGG2010}, and \cite{ait2015modeling}.

Since Hawkes processes can be easily incorporated with a classical actuarial frequency model for systemic cyber risk, they can be integrated into the standard collective risk model if complemented by an appropriate severity modeling approach.  In principle, the severities of systemic events could be chosen as described in Section \ref{subsec:classicalseverity} for idiosyncratic and systematic events. Due to the limited amount of data and uncertainty about the possible impact of future systemic cyber incidents, accurate modeling of systemic severities is extremely challenging in practice.

Hawkes processes take a top-down approach to modeling systemic cyber risk and neglect the specific infection processes that underlie risk contagion in interconnected systems. Important aspects of risk amplification and possible accumulation scenarios may not be adequately captured. This is the main attractive feature of epidemic network models; their disadvantage is their increased complexity.
	
\subsection{Epidemic Network Models}\label{subsec:interconnectednetworkmodels}
Interconnectedness constitutes a key characteristic of cyber systems. Systemic cyber risks may spread and amplify in networks of interconnected companies, economic actors, or financial institutions.
Cyber network models for contagious risk propagation consist of the following three key components:
\begin{enumerate}
\item A \textbf{network} (also called \emph{graph}) whose nodes represent components or agents. These entities could be individual corporations, subsystems of computers, or single devices. The edges of the network correspond to possible transition channels, e.g., IT connections or exchange of data/computer code, see Section \ref{subsubsec:networks};
\item A \textbf{spread process} on the network that models the propagation of a contagious cyber risk, like the spread of a
computer virus, a Trojan, or ransomware,\footnote{ Moreover, contagion can also be interpreted in a broader sense, e.g., considering the propagation of business interruptions or the breakdown of supply chains as the consequence of cyber attacks on single entities.}  see Section \ref{subsubsec:spreadprocess};
\item  A \textbf{loss model} which determines the severity of cyber events and the monetary impact on different agents in the network, see Section \ref{subsubsec:lossmodel}.
\end{enumerate}

\subsubsection{Networks}\label{subsubsec:networks}

\begin{definition}[Network]
A \( \emph{network}\)\footnote{This definition refers to unweighted networks. In the context of weighted networks, the notion of undirected networks refers to the symmetry of the weight matrices.} (or \emph{graph}) \(G\) is an ordered pair of sets \( G= (\mathcal{V}, \mathcal{E} )\), where \(\mathcal{V} \neq \emptyset \) is a countable set of $N$ elements, called \(\emph{nodes}\) (or \(\emph{vertices}\)), and \(\mathcal{E}\) is a set of pairs \( (i, j)\), \(i, j \in \mathcal{V}\), of different nodes, called \(\emph{edges}\) (or \(\emph{links}\)).
If all edges in \(\mathcal{E}\) are unordered, formally, $(i,j) \in \mathcal{E}  \Rightarrow (j,i) \in \mathcal{E}$, then \(G\) is called an \( \emph{undirected network}\). Otherwise, the network \(G\) is called \( \emph{directed}\). \\
\end{definition}

The network structure is encoded in its \emph{adjacency matrix} $A=(a_{ij})_{i,j\in\{1,\ldots,N\}}\in\{0,1\}^{N\times N}$, which is defined by its entries
\begin{equation*}
a_{ij} :=
\begin{cases}
1, &  \textmd{if}  \textbf{ }(i, j) \in \mathcal{E}\\
0, &  \textmd{if} \textbf{ }(i, j) \notin \mathcal{E}.
\end{cases}
\end{equation*} 
By definition, $G$ is undirected if and only if \( A\) is symmetric. Examples of undirected network topologies with $N=8$ nodes are depicted in Figure \ref{fig: networks}.

\begin{figure}[h]
	\begin{center}
		\begin{minipage}[t]{0.23\linewidth}
			\centering
			\begin{tikzpicture}[scale=1.2] 
				\tikzstyle{every node}=[draw,fill=White!85!gray, shape=circle];
				\node (1) at (0, 1) {1};
				\node (2) at ( 0.707, 0.707) {2};
				\node (3) at (1, 0) {3};
				\node (4) at (0.707, -0.707) {4};
				\node (5) at (0, -1) {5};
				\node (6) at (-0.707, -0.707) {6};
				\node (7) at (-1, 0) {7};
				\node (8) at (-0.707, 0.707) {8};
			\end{tikzpicture}
			\caption*{isolated nodes}
		\end{minipage}
		\begin{minipage}[t]{0.23\linewidth}
			\centering
			\begin{tikzpicture}[scale=1.2] 
				\tikzstyle{every node}=[draw,fill=White!85!gray, shape=circle];
			\node (1) at (0, 0) {1};
            \node (2) at ( 0, 1) {2};
            \node (3) at (0.78183, 0.62349) {3};
            \node (4) at (0.97493, -0.2225) {4};
            \node (5) at (0.43388, -0.901) {5};
            \node (6) at (-0.43388, -0.901) {6};
            \node (7) at (-0.97493, -0.2225) {7};
            \node (8) at (-0.78183, 0.62349) {8};

            \draw 
            (1) -- (2)
            (1) -- (3)
            (1) -- (4)
            (1) -- (5)
            (1) -- (6)
            (1) -- (7)
            (1) -- (8);
			\end{tikzpicture}
			\caption*{star-shaped}
		\end{minipage}
		\begin{minipage}[t]{0.23\linewidth}
			\centering
			\begin{tikzpicture}[scale=1.2] 
				\tikzstyle{every node}=[draw,fill=White!85!gray, shape=circle];
				\node (1) at (0, 1) {1};
				\node (2) at ( 0.707, 0.707) {2};
				\node (3) at (1, 0) {3};
				\node (4) at (0.707, -0.707) {4};
				\node (5) at (0, -1) {5};
				\node (6) at (-0.707, -0.707) {6};
				\node (7) at (-1, 0) {7};
				\node (8) at (-0.707, 0.707) {8};
				\draw (1) -- (2)
				(1) -- (3)
				(1) -- (4)
				(1) -- (5)
				(1) -- (6)
				(1) -- (7)
				(1) -- (8)
				(2) -- (3)
				(2) -- (4)
				(2) -- (5)
				(2) -- (6)
				(2) -- (7)
				(2) -- (8)
				(3) -- (4)
				(3) -- (5)
				(3) -- (6)
				(3) -- (7)
				(3) -- (8)
				(4) -- (5)
				(4) -- (6)
				(4) -- (7)
				(4) -- (8)
				(5) -- (6)
				(5) -- (7)
				(5) -- (8)
				(6) -- (7)
				(6) -- (8)
				(7) -- (8);
			\end{tikzpicture}
			\caption*{fully connected}
		\end{minipage}
		\begin{minipage}[t]{0.23\linewidth}
			\centering
			\begin{tikzpicture}[scale=1.2] 
				\tikzstyle{every node}=[draw,fill=White!85!gray, shape=circle];
				\node (1) at (-2, 0) {1};
               \node (2) at ( -1, 0) {2};
              \node (3) at (-0.293, 0.707) {3};
             \node (4) at (-0.293, -0.707) {4};
                \node (5) at (0.647, 1.049) {5};
                \node (6) at (0.647, 0.365) {6};
                \node (7) at (0.647, -0.365) {7};
                \node (8) at (0.647, -1.049) {8};
                \draw 	(1) -- (2)
                        (2) -- (3)
                        (2) -- (4)
                        (3) -- (5)
                        (3) -- (6)
                        (4) -- (7)
                        (4) -- (8);
			\end{tikzpicture}
			\caption*{branching tree}
		\end{minipage}
	\end{center}
	\caption{Examples of network topologies with $N=8$ nodes.}
	\label{fig: networks}
\end{figure}

In applied network analysis, the exact network structure is often unknown. In this case, {random network models} enable sampling from a class of networks with given fixed topological characteristics (such as the overall number of nodes).\footnote{Two commonly used models are discussed in Appendix \ref{app:randomnet}.}

In the cyber insurance literature, network models are mainly applied to study risk contagion, e.g., modeling the propagation of malware in IT networks of interconnected firms or devices. In addition to an underlying network, an appropriate model of the contagion process that captures epidemic spread is needed.
	
\subsubsection{Epidemic Spread Processes}\label{subsubsec:spreadprocess}
	Models of infectious disease spread dynamics have been studied extensively in mathematical biology and epidemiology, dating back at least to the seminal work of \cite{Kermack1927}.\footnote{The models typically focus either on an epidemic spread within a population, as, e.g., in \cite{Kermack1927}, or on the spread along paths of a predefined network; for a detailed overview, see, e.g., \cite{PastorSatorras2015} and \cite{Kiss2017}.} In this paper, we focus on epidemic \emph{network} models for populations of entities.
	
	At any point in time, each node is in a particular state, which may change over time as it interacts with other nodes. According to their state, individuals are divided into various \textit{compartments}, e.g., individuals that are \emph{susceptible} $(S)$ to an infection, \emph{infected} $(I)$ individuals, or individuals who have \emph{recovered} $(R)$ from the infection. For a network of $N$ nodes, the spread process at time $t$ can be described by a \textit{state vector} 
 \begin{equation*}
 X(t) = (X_1(t),\ldots , X_N(t)) \in E^N,
 \end{equation*}
 where $E$ is the set of compartments. Both \textit{Markov} and \textit{non-Markov} processes have been considered in the context of epidemic spread processes.\footnote{The Markov property captures that a process is "memoryless", i.e., that the conditional distribution of future values $X_{t+s}$, $s>0$, of the process does only depend on the present value of the process $X_t$ and not additionally on past values $X_\mu$, $\mu < t$.}
	
\paragraph{Markovian Spread Models}
In Markovian spread models on networks, the evolution of the state vector $X(t)$ is described by a (in many cases: time-homogeneous) continuous-time Markov chain on the discrete state space $E^N$. The Markov models SIS \textit{(Susceptible-Infected-Susceptible)} and SIR \textit{(Susceptible-Infected-Recovered)} form a class of commonly used models for epidemic propagation in networks. They are distinguished by the presence (SIR) or absence (SIS) of immunity: Reinfection events are only possible in the SIS framework because in the SIR model recovered individuals acquire (permanent) immunity, i.e., the models are based on the two different sets of compartments $E=\{S,I\}$ and $E=\{S,I,R\}$.

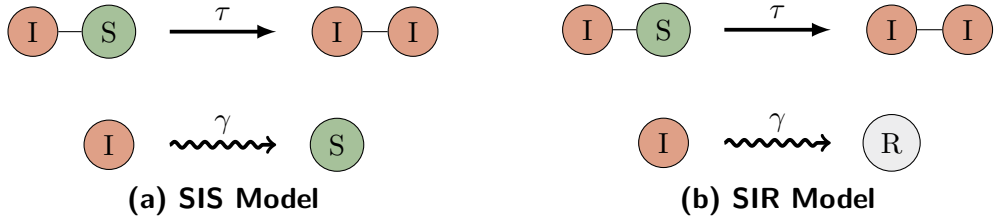
\begin{figure}[h]
		\centering
		\begin{minipage}[t]{0.45\linewidth}
				\centering
				\begin{tikzpicture}[scale=1] 
						\node[draw, shape=circle] (1)[fill=BrickRed!40!White] at (0, 0) {I};
						\node[draw, shape=circle] (2) [fill=OliveGreen!40!White] at ( 1, 0) {S};
						\node[draw, shape=circle] (3)[fill=BrickRed!40!White] at (4, 0) {I};
						\node [draw, shape=circle](4) [fill=BrickRed!40!White] at ( 5, 0) {I};
						\draw (1) -- (2)
						(3) -- (4);
						\draw[arrow, line width=0.5mm] (1.8,0) --  node[anchor=south] {$\tau$} (3.2,0) ;
						\node[draw, shape=circle] (5) [fill=BrickRed!40!White] at (1,-1.5) {I};
						\node[draw, shape=circle](6) [fill=OliveGreen!40!White] at (4,-1.5) {S};
						\draw [->,decorate,decoration={snake,amplitude=.4mm,segment length=2mm,post length=1mm}, line width=0.5mm] (1.8,-1.5) -- node[anchor=south] {$\gamma$}(3.2,-1.5);
					\end{tikzpicture}\\
				\textbf{\sffamily (a) SIS Model}
			\end{minipage}
		\begin{minipage}[t]{0.45\linewidth}
				\centering
				\begin{tikzpicture}[scale=1] 
						\node[draw, shape=circle] (1)[fill=BrickRed!40!White] at (0, 0) {I};
						\node[draw, shape=circle] (2) [fill=OliveGreen!40!White] at ( 1, 0) {S};
						\node[draw, shape=circle] (3)[fill=BrickRed!40!White] at (4, 0) {I};
						\node [draw, shape=circle](4) [fill=BrickRed!40!White] at ( 5, 0) {I};
						\draw (1) -- (2)
						(3) -- (4);
						\draw[arrow, line width=0.5mm] (1.8,0) --  node[anchor=south] {$\tau$} (3.2,0) ;
						\node[draw, shape=circle] (5) [fill=BrickRed!40!White] at (1,-1.5) {I};
						\node[draw, shape=circle](6) [fill=White!85!gray] at (4,-1.5) {R};
						\draw [->,decorate,decoration={snake,amplitude=.4mm,segment length=2mm,post length=1mm}, line width=0.5mm] (1.8,-1.5) -- node[anchor=south] {$\gamma$}(3.2,-1.5);
					\end{tikzpicture}\\
				\textbf{\sffamily (b) SIR Model}
			\end{minipage}
		\caption{Infection and recovery for the SIS and SIR network model.}
		\label{fig:SISSIR}
	\end{figure}

In both models, a transition of $X$ from one state in $E^N$ to another is possible only if exactly one node changes its state $X_i$ in $E$. State changes may occur through infection or recovery: It is assumed that each node may be infected by its infected neighbors, but can be cured independently of all other nodes in the network. Each node is endowed with an independent exponential clock and changes its state when the exponential clock rings. Letting $\tau>0$ and $\gamma> 0$, the rates of these transitions are illustrated in Figure \ref{fig:SISSIR} and given as follows $(i=1,\ldots,N)$:
    \begin{align}
    \begin{split}\label{eq:SISSIRrates}
        X_i: S\rightarrow I\quad  &\text{with rate}\quad \tau \sum_{j=1}^N a_{ij} \mathbbm{1}_{\{X_j(t) =I\}} \\
         X_i: I\rightarrow Z\quad  &\,\text{with rate}\quad \gamma,
    \end{split}
    \end{align}

where $Z=S$, for the SIS, and $Z=R$ for the SIR model, respectively.

The exponential transition times enable an intuitive stochastic simulation algorithm: the well-known \textit{Gillespie algorithm}, first introduced in \cite{Gillespie1976} and \cite{Gillespie1977}; see Appendix \ref{app:Gillespie} for details.

For practical purposes such as the pricing of cyber insurance contracts, we often do not need the full information provided by the Markov chain evolution, but only the dynamics of specific quantities such as moments or (infection) probabilities. Of particular interest are the dynamics of the state probabilities of individual nodes $\mathbb{P}(X_i(t)=x_i),\; t\ge 0$.
They can be derived from Kolmogorov's forward equation and written in general form as ($i=1,\ldots,N$)
\begin{equation}\label{eq: master}
	\frac{d\mathbb{P}(X_i(t) = x_i)}{dt} = \sum_{y: y_i=x_i} \sum_{z\neq y} [\mathbb{P}(X(t)={z}) q_{zy} - \mathbb{P}(X(t)={y}) q_{yz}],
\end{equation}
where $q_{zy}$ denotes the transition rate of the entire process $X$ from $z\to y$.
In natural sciences, this equation is also known under the term \emph{master equation}. For the SIS and SIR models, using Bernoulli random variables $S_i(t):= \mathbbm{1}_{\{X_i(t) = S\}}$, $I_i(t):= \mathbbm{1}_{\{X_i(t) = I\}}$, and (for SIR) $R_i(t):= \mathbbm{1}_{\{X_i(t) = R\}}$, the dynamics of state probabilities of individual nodes \eqref{eq: master} can conveniently be written via moments:

\begin{itemize}
		\item {\textbf{SIS model:}}\footnote{In the cyber insurance literature, the SIS Markov model was used by \cite{fahrenwaldt2018pricing}.
				Also, a brief application was studied in \cite{xu2019cybersecurity} with a modified $\varepsilon$-SIS model, originally proposed in \cite{Mieghem2012}. Here, an infectious threat for node $i$  from outside the network is included with a rate $\varepsilon_i$.}
Since $E =\{ I,S\}$, we have $S_i(t)=1-I_i(t)$, i.e., the evolution of $X$ is fully described by the evolution of the vector $I(t)=(I_1(t),\ldots , I_N(t))$, and the single node infection dynamics are given by
\begin{equation}\label{eq:SIS}
\frac{d \mathbb{E}[I_i(t)]}{d t} = -\gamma \mathbb{E}[I_i(t)] + \tau \sum_{j=1}^N a_{ij}\mathbb{E}[I_j(t)] - \tau \sum_{j=1}^N a_{ij}\mathbb{E}[I_i(t)I_j(t)],\quad i=1,\ldots , N,
\end{equation}
since $\mathbb{P}(X_i(t)=I)=\mathbb{P}(I_i(t)=1)=\mathbb{E}[I_i(t))].$ This system of $N$ equations is not closed as second order moments $\mathbb{E}[I_i(t)I_j(t)]$, i.e., second order infection probabilities, appear.

\item {\textbf{SIR model:}} The dynamics of the recovery Bernoulli random variable $R_i(t)$ result from the dynamics of $I_i(t)$ and $S_i(t)$ due to $\mathbb{E}[R_i(t)] = 1-\mathbb{E}[S_i(t)] - \mathbb{E}[I_i(t)]$. Equation \eqref{eq: master} corresponds to:
\begin{align}\label{eq:SIR}
\begin{split}
		\frac{d\mathbb{E}[S_i(t)]}{dt} &= -\tau\sum_{j=1}^N a_{ij} \mathbb{E}[S_i(t)I_j(t)] , \\
		\frac{d\mathbb{E}[I_i(t)]}{dt} &= \tau\sum_{j=1}^N a_{ij} \mathbb{E}[S_i(t)I_j(t)] - \gamma \mathbb{E}[I_i(t)] ,\\
	\end{split}
\end{align}
for $i = 1,2,\ldots , N$. Again, the system is {not closed} due to the presence of second order moments.
\end{itemize}

The main problem with systems \eqref{eq:SIS} and \eqref{eq:SIR} is the fact that they are \textit{not closed}: They depend on second order moments, which, in turn, depend on third order moments, etc. For example, the fully closed SIS model yields $\sum_{i=1}^N \binom{N}{i} = 2^N-1$ moment (i.e., infection probability) equations. Solving these systems exactly becomes intractable for networks of realistic size. To deal with this issue, the following two approximation approaches have been proposed:
\begin{enumerate}
\item \textbf{Monte Carlo simulation:}  Monte Carlo simulation using the Gillespie algorithm (Algorithm \ref{alg:Gillespie} in the appendix) constitutes a powerful tool to obtain various quantity estimates related to the evolution of the epidemic spread. In particular, this includes the state probability dynamics of individual nodes \eqref{eq: master}.\footnote{
Pseudocode and further explanations of the Gillespie algorithm applied to the SIS and SIR epidemic network models is, e.g., given in Appendix A.1.1 of \cite{Kiss2017}.}
\item \textbf{Moment closures:} 
If a set of nodes $J\subset\mathcal{V}$ is infected, this increases the probability of other nodes in the network (that are connected to the set $J$ via an existing path) to become infected as well. Node states do not evolve independently and are to some extent \textit{correlated}.
To break the cascade of equations and to make ODE systems tractable, the moment closure approach consists in factorizing moments beyond a certain order $k$, substituting all higher-order moments. This is done by considering the exact moment equations up to this order $k$ and \textit{closing} the system by approximating moments of order $k+1$ in terms of products of lower-order moments using a mean-field function. 
A detailed description of two different types of moment closures is provided in Appendix \ref{app:momentclosures}. 
However, a major problem with moment closures is that only little is known about rigorous error estimates.\footnote{This problem has also been highlighted in the epidemic literature, see, e.g., \cite{Kiss2017}, p.115.} This presents an important avenue for future research.
\end{enumerate}

\paragraph{Non-Markovian Spread Models}
Non-Markovian models possess conditional distributions that may depend on the past and on further random factors. In contrast to the Markovian setup, where transition times are necessarily exponential, non-Markovian models might allow additional flexibility to freely choose the distributions of infection and recovery times. In addition, dependence among the infection times may be included. This generality may improve the quality of a fit to real-world data. However, the extended generality in comparison to Markov models is typically associated with reduced tractability. For this reason, non-Markovian models are less commonly considered. In addition, a similar scope of flexibility can also be achieved within the class of Markovian models by extending the dimension of the state space; but this comes again at the price of increased complexity and possibly reduced tractability.

A simple example of a non-Markovian model for the spread of cyber risks has been proposed by \cite{xu2019cybersecurity}. The model does not include immunity, i.e., the underlying compartment set is the same as for the Markovian SIS model. The considered waiting times in the model are:
	\begin{itemize}
	    \item The individual \textbf{recovery times} $T^{recov}_i$ of \textit{infected} nodes.
	    \item
	    For nodes $i$ which are in the \textit{susceptible} state, two different types of infections are considered, \textit{internal infections} from within the network and \textit{external infections} coming from outside:
	    \begin{enumerate}
	        \item \textbf{Internal infection times:}
	        Let the random variable $K_i(t) = \sum_{j=1}^N a_{ij} I_j(t)$ denote the number of infected neighbors of node $i$ at time $t$. Infectious transmissions to node $i$ are given with waiting times $T_{i_1},\ldots ,T_{i_{K_i}}$. These times share the same marginal distribution $F_i$. Their underlying dependence structure is captured by a prespecified copula.
	        \item \textbf{External infection times:} A random variable $T^{out}_i$ with distribution $G_i$ models the arrival time of threats from outside the network to node $i$. $T^{out}_i$ is assumed to be independent of times $T_{i_1},\ldots ,T_{i_{K_i}}$.
	    \end{enumerate}
\end{itemize}
To simulate the process, the waiting times for all nodes are generated according to their current state (i.e., recovery times for all infected nodes, and internal and external infection times for all susceptible nodes). The minimum of these waiting times determines the next event (infection or recovery). After this change, all quantities are recomputed and the process is repeated until a prespecified stopping criterion is met.\footnote{Pseudocode for stochastic simulations is provided in Algorithm 1 of \cite{xu2019cybersecurity}.} 

Finally, note that a Markovian SIS model with outside infections\footnote{To be precise, the so-called $\varepsilon$-SIS model, originally proposed in \cite{Mieghem2012}, arises.} can be obtained as a special case by choosing exponentially distributed infection and recovery times and assuming independence between all waiting times.

\subsubsection{Loss Models}\label{subsubsec:lossmodel}
Given the underlying network, and the epidemic spread process $X$ on it, the third and final ingredient of a cyber risk network model is given by a suitable loss model $Y_{i,j}$ for each node $i=1,\ldots,N$, where $j$ describes the number of loss events. In the existing literature, loss models are kept rather simple as the focus lies on modeling the cyber-epidemic spread. We give two examples:
  \begin{enumerate}
      \item In \cite{fahrenwaldt2018pricing}, cyber attacks are launched in a two-step procedure: First, using a random process, times of attacks on the entire network (loss events) $t_1, t_2,\ldots$ are generated. Second, for each node $i$, a possible random loss $L_{i,j}$ is modeled, where $j$ describes the index of the corresponding attack time. The loss, however, only materializes if node $i$ is infected at the attack time. This is captured by the loss model
     \begin{equation*}
        Y_{i,j} = L_{i,j} \cdot \mathbbm{1}_{X_i(t_j)=I}, \quad i=1,\ldots N, \quad j =1,2, \ldots.
     \end{equation*}
      \item In \cite{xu2019cybersecurity}, the loss model $Y_{i,j}$ is given by
      \begin{equation*}
          Y_{i,j} = \eta_i (D_{i,j}) + C_i(T^{recov}_{i,j}), \quad i=1,\ldots N, \quad j = 1,2,\ldots
      \end{equation*}
      with a legal cost function $\eta_i$, the number $D_{i,j}$ of data damaged in the infection $j$, 
      and  a  cost function $C_i$ depending on the recovery time $T^{recov}_{i,j}$ of node $i$ for infection event $j$. Here, 
      the recovery time $T^{recov}_{i,j}$ for each event $j$ is obtained from the infection dynamics,
      while the data loss sizes $D_{i,j}$ are assumed to follow a beta distribution.
  \end{enumerate}

\subsubsection{On Calibration and Application}
Epidemic network models in the cyber insurance literature mostly focus on a general assessment of the underlying structure of systemic cyber risks: aspects of risk contagion and propagation are characterized in a \textit{qualitative} sense. For example, \cite{fahrenwaldt2018pricing} study the effect of homogeneous, star-shaped, and clustered  topologies on the resulting overall insurance losses in regular networks, demonstrating the strong impact of the network topology on risk propagation. Further, epidemic network models could also be applied to identify critical initial infection locations or critical network links that may augment cyber losses. The models are thus particularly useful for counterfactual simulations and have not yet been calibrated to real-world data.

More applications of epidemic network models to cyber risk contagion can be found in the engineering literature. However, these works do not study risk emergence on a global level. Instead, they analyze cyber risks which are building from the microstructure of interconnected IT devices in \textit{local} environments. For example, \cite{Powell2020} focuses on local IT authentication procedures, where the corresponding vectors of lateral movements within a network can be interpreted as edges of a directed mathematical graph. Possible attack vectors are evaluated using classical metrics from network theory and epidemic spreading models of SIR type. More technical and IT-related aspects of cyber security issues in smart grids, i.e., networked power systems for energy production, distribution, and consumption, are surveyed and discussed in \cite{Wang2013}.

However, for the \emph{quantitative} assessment of systemic cyber risk from a regulatory or actuarial perspective, contagion among different companies needs to be studied on a \textit{global} scale. A major challenge for accurate modeling is the estimation of the exact network structure and the epidemic parameters of past and future incidents -- particularly due to data limitations and the speed of technological evolution. In Appendix \ref{app:NetworkEstimation}, we provide a brief overview and classification of existing estimation approaches for epidemic network models that are not necessarily related to cyber; in our view, however, it is conceivable, that such approaches could also be implemented and further developed in a cyber context in future cyber risk research.

To overcome the estimation challenge, \textit{top-down approaches} have been proposed in the literature. In \cite{Hillairet2021}, the impact of massive global-scale cyber-incidents, like the WannaCry scenario, on insurance losses and assistance services is determined. While network contagion is implicitly considered, it is not modeled within an actual network framework; instead, the authors choose the original population-based SIR model of \cite{Kermack1927} which describes {deterministic} dynamics of the {total numbers} of susceptible, infected, and recovered individuals within the global population of IT devices. The corresponding ODE system is given by
{\begin{align*}
    \frac{d S(t)}{dt} &= -\tau S(t) I(t) \\
    \frac{d I(t)}{dt} &= \tau S(t) I(t) - \gamma I(t) \\
    \frac{d R(t)}{dt} &= \gamma I(t)
\end{align*}}with constant global population size $N = S(t) + I(t) + R(t)$. Parameters $N$, $\tau$, and $\gamma$ are estimated from data of the WannaCry cyber incident. 

Given this global spread, the focus of the paper lies on the \textit{stochastic} evolution of the insurer's \emph{local} portfolio consisting of $n<<N$ policyholders and their corresponding losses. The influence of the global cyber epidemic on the local portfolio is captured by the hazard rate $\lambda_{T_i^{infec}}$ of the policyholders' infection times $T_i^{infec}$:
{\begin{equation*}
    \lambda_{T_i^{infec}}(t) := \lim_{dt\to 0^+}\frac{1}{dt} \mathbb{P}(T_i^{infec}\in [t, t + dt]\mid T_i^{infec}\geq t) = \tau I(t),
\end{equation*}}
i.e., the local hazard rates are assumed to be proportional to the number of infected individuals in the global population.

Most recently, this model has further been extended by replacing the homogeneous global population model with a network scenario of interconnected industry sectors, see \cite{Hillairet2021b}. The underlying {directed} and {weighted} network structure is derived from OECD data that measures the economic flow between different industries, and this data is interpreted as a reasonable estimate of the digital dependence between these sectors. Contagion between sectors is modeled using a deterministic multi-group SIR model for the total numbers of susceptible, infected, and recovered companies in the sectors. Due to the scarcity of data currently available, such top-down approaches present promising avenues for risk management and actuarial modeling.

Additionally, future research should analyze the implementation of more realistic loss models, that, e.g., contain different types of cyber events and capture their characteristic severity distributions (see also the discussion on classical frequency-severity approaches in Section \ref{subsec:classicalseverity}). This would further strengthen the applicability of network models in practice.

	
	\subsection{Game-Theoretic Models and Strategic Interaction Effects}\label{subsec:strategicinteraction}
	
In addition to contagion due to the interconnectedness of entities in cyber networks, potentially different objectives of the actors and their strategic interaction constitute a key characteristic of systemic cyber risk. The risk exposure of individuals is often interdependent, since it is influenced by the behavior of other actors.  \textit{Game theory} provides a suitable framework to study this dimension in the cyber ecosystem. 

In the first part of this section, we  briefly review and provide a short mathematical introduction to game theoretic approaches applied to study cyber risk and cyber insurance (Section \ref{sec:gametheorymodels}). For an exhaustive review of the corresponding literature, we refer to the  surveys \cite{Boehme2010}, \cite{Boehme2018}, and \cite{Marotta2017}. We will adopt the notation from \cite{Marotta2017}.
Section~\ref{sec:critics} evaluates the considered models.

\subsubsection{Game-Theoretic Modeling Approaches}\label{sec:gametheorymodels}

The majority of game theoretic contributions focuses on {self protection of interdependent actors} in the presence or the absence of cyber insurance. A key question is whether and under which conditions cyber insurance provides incentives for self protection and improves global IT security. In this section, we present\footnote{We refer to \cite{Marotta2017} for an in-depth overview of the topic.} the main ideas and characteristics of such models.

\paragraph{Three Different Types of Actors in the Game}
 
We consider three types of strategic players 
with different objectives: potential buyers of insurance (for simplicity, called agents), insurance companies, and the regulator.
\begin{enumerate}
\item \textbf{Agents} are the potential cyber insurance policyholders. To capture interdependence, most models assume that agents form a network. Agent $i$ aims to maximize her expected utility
\begin{equation*}
	\max \mathbb{E}[U_i(W_i)],
	\end{equation*}
where 
\begin{itemize}
    \item $U_i$ denotes the utility function of agent $i$. Various types of utility functions are considered in the literature; most of them satisfy the classical von-Neumann-Morgenstern axioms. While some papers, such as \cite{liu2014}, \cite{Pal2012}, and \cite{Pal2014}, allow for {\em heterogeneous preferences}, the majority of models assumes {\em homogeneous preferences}, i.e., $U_i=U$ across all agents. 
    \item $W_i$ is the financial position of agent $i$ at the end of the insurance period.  The value $W_i$ depends on whether the agent has bought an insurance contract or not, on her investment $C_i$ in cyber security, and on potential losses $L_i$ in case the agent is affected by a cyber attack.
\end{itemize} 
 
The agent's \textbf{self protection level} $x_i$ is a \textit{crucial model component} when studying interdependence.\footnote{Only few papers, e.g. \cite{bohme2005cyber}, \cite{bohme2006models}, \cite{Johnson2014} and \cite{Johnson2014a}, do not include self protection in the model.}  Most of the existing literature falls into either of the following two distinct categories: Some assume that only two security states are possible, secured or not, with the corresponding constant cost $C$ or $0$. Others propose a continuous scale of security levels, e.g., $x_i\in[0,1]$. The value of $x_i$ affects
\begin{itemize}
    \item \emph{the cost of self protection $C_i$}: \\ For a continuous spectrum of security levels, i.e., $x_i\in [0,1]$, $C_i=C(x_i)$ is typically assumed to be an increasing convex function of $x_i$, reflecting that user costs rapidly increase when  improving security.
    \item \emph{agent $i$'s probability of becoming infected $p_i := \mathbb{P}(I_i=1)$}: \\ Obviously, this probability depends on the individual security level $x_i$ of the agent $i$, but -- due to interdependence -- it may also be influenced by the individual security levels of {other} network participants.
\end{itemize}

Within this framework, agent $i$'s expected utility can be computed
\begin{enumerate}
    \item \textbf{without insurance:} 
    \[\mathbb{E}[U_i(W_i)] = (1-p_i)\cdot U_i(W_i^0-C_i) + p_i\cdot U_i(W_i^0-L_i-C_i)\]
    \item \textbf{with insurance:} 
    \[\mathbb{E}[U_i(W_i)] = (1-p_i)\cdot U_i(W_i^0-\pi_i-C_i) + p_i\cdot U_i(W_i^0-L_i-C_i-\pi_i+\hat{L}_i)\]
\end{enumerate}

where
\begin{itemize}
\item $W_i^0$ denotes the initial wealth of agent $i$.
\item $\pi_i$ is the insurance premium of agent $i$ set by the insurer. This premium depends on the type of insurance market; we will discuss different models below. 
\item $L_i$ is the potential loss of agent $i$ that is governed by a binary distribution: only two possible scenarios are considered. Either the agent experiences a cyber attack with a \textit{fixed loss size}, or she is not attacked which corresponds to no loss. This particular setting excludes the possibility of different types of cyber attacks. Multiple attacks are also not considered.\footnote{We will discuss the scope of the existing models in Section~\ref{sec:critics}.}
The majority of game theoretic models relies on the assumption of constant homogeneous losses for all agents, i.e., $L_i\equiv L$. 
\item $\hat{L}_i$ is the cover in case of loss which is specified in the insurance contract. Most papers assume full coverage, i.e., $\hat{L}_i=L_i$, but some consider alternatively partial coverage, e.g., in order to mitigate the impact of information asymmetries, cf.~\cite{Mazzoccoli2020}, \cite{Pal2012}, \cite{Pal2014}.
	\end{itemize}

\item \textbf{Insurance Companies}: The insurer sets the cyber insurance premiums and specifies the insurance cover $\hat{L}_i$. Insurance premiums depend on the market structure:
\begin{itemize}
	\item \textbf{Competitive market}: This is the prevailing model in the literature. The profits of the insurers are zero in this case; customers pay fair premiums. Competitive markets are a boundary case that almost surely leads to the insurer's ruin in the long run.
	\item \textbf{Monopolistic market / Representative insurer}: Another extreme is a market with only one insurance company. In these models, the impact of a monopoly can be studied. An alternative consists in studying objective functions that are different from the insurer's profit. This situation is mostly studied in the context of regulation: The insurer represents a regulatory authority and is not aiming for profit maximization, but focuses on the wealth distribution in order to incentivize a certain standard of IT protection.\footnote{Market models of this type are studied in \cite{liu2014} with a zero-profit insurer. Profits are still possible in \cite{Pal2012}, \cite{Pal2014} and \cite{Pal2019}, and maximized in \cite{Khalili2017}.}
\item \textbf{Immature market/Oligopoly}: Instead of a monopoly, imperfect competition is studied with multiple insurers that may earn profits. The increments between the fair price and the insurance premium is determined by the markets structure.\footnote{Immature markets are considered, e.g., in \cite{Martinelli2017}, \cite{Martinelli2016}, \cite{Ogut2005}.} 
	\end{itemize}

\item \textbf{Regulator:} 
Market inefficiencies and a lack of cyber security may be mitigated by regulatory policies.  \textit{Regulatory instruments} include mandatory insurance, fines and rebates, liability for contagion, etc. The choice of policies and their impact can be studied\footnote{The effects of such regulatory instruments were, e.g., studied in \cite{Bolot2009}, \cite{liu2014}, \cite{Pal2012}, \cite{Pal2014}.} by introducing a third party, the regulator. The objective of the regulator is to maximize a \textit{social welfare function}. This could, for example, be chosen as the sum of the expected utilities of the agents
	\begin{equation*}
	\sum_i \mathbb{E}[U_i(W_i)].
	\end{equation*}
\end{enumerate}

\paragraph{Interdependent Self Protection in IT Networks}\label{sec:cyber_characteristics1}

The strategic interaction of the three types of players introduced above is modeled as a game.   The agents form an interconnected network and optimize their expected utility. Their individual security level and the amount of cyber insurance coverage serve as their controls. The insurance companies are provider of risk management solutions. In some models, a regulator is included as a third party with the aim to improve welfare, e.g., by implementing standards of protection in cyber systems.

The network topologies are, typically, quite stylized to guarantee tractability. For example, two-agent models are considered in \cite{Ogut2005}. Most papers investigate complete graphs, e.g., \cite{Ogut2005}, \cite{schwartz2014cyber} and \cite{Pal2014}. \cite{Bolot2009} and \cite{yang2014}, in contrast, investigate networks with degree heterogeneity, but restrict their analysis to Erdős-R\'enyi random graphs.

Agents are interdependent in the network, since the infection probability $p_i$ depends on the local security level $x_i$ and levels of the other nodes $y_{i} := (x_1,\ldots , x_{i-1}, x_{i+1}, \ldots , x_N)$ (or at least of $i$'s neighbors). In some cases, $p_i$ is assumed to depend on an overall network security level as well.\footnote{This is the case in \cite{shetty2010}, \cite{shettyschwartz2010} and \cite{schwartz2014cyber}.} However, in contrast to the models from Section \ref{subsec:interconnectednetworkmodels}, attacks do not result from a dynamic contagion process; instead, the infection is assumed to be \textit{static} and the values $p_i$ are derived from \textit{ad hoc schemes}. The most common one\footnote{An alternative approach using a simplified two-state scenario of security investments is analyzed in \cite{Lelarge2008}, \cite{Lelarge2008a}, and \cite{yang2014}. Infection probabilities are derived from a recursive branching process.} 
assumes a continuous spectrum of security levels and computes $p_i$ as the complementary probability of the case that neither a direct nor an indirect attack occurs:
\begin{equation*}
\label{eq:prob}
\begin{split}
p_i (x_i, y_{i})&= 1-(1-p_i^{dir})(1-p_i^{cont}) \\
&= 1-(1-\psi_i(x_i))\times \prod_{j\neq i}(1-h_{i,j}\psi_j(x_j))
\end{split}
\end{equation*}
where
\begin{itemize}
\item $p_i^{dir} = \psi_i(x_i)$ denotes the probability of direct infection of $i$ through threats from outside the network. It is interpreted as a function of the individual security level $x_i$.
\item $p_i^{cont} = 1-\prod_{j\neq i} (1-h_{i,j}\psi_j(x_j))$ is the probability for node $i$ to become infected through contagion. The probability for $i$ to be infected via node $j$ is given by $h_{i,j}$, i.e., $h_{i,j}\neq 0$ only if $i$ and $j$ are adjacent. This is where the underlying network topology comes into play.
\end{itemize}

In the absence of information asymmetries, the game between agents and the insurer(s) involves three perspectives:\footnote{Variations of the game design are possible; e.g., in \cite{Laszka2018} the authors use a signaling game instead of a game model to study the adverse selection problem, allowing insurers to audit the agents' security. A similar game is considered in \cite{Khalili2017} who introduce a pre-screening procedure.}  
\begin{enumerate}
\item A legal framework is set by the regulator (if a regulator is present).
\item Agents specify their levels of self protection and insurance protection and select from the available contract types to maximize their expected profits.
\item Insurance companies compute the corresponding contract details, i.e.,  premiums $\pi_i$ and indemnities $\hat{L}_i$. In absence of information asymmetries between agents and the insurer(s), the protection levels of policyholders can be observed by the insurer and are reflected by the contract.
\end{enumerate}
The model may be augmented to incorporate information asymmetries:
\begin{itemize}
    \item \textbf{Moral hazard}: A dishonest policyholder may behave in a way that increases the risk, if the insurer cannot properly monitor the policyholder's behavior. In the game, this is represented by the possibility for agents to modify their self protection\footnote{Some authors distinguish between \emph{self-insurance} -- a reduction in the size of a loss -- and \emph{self-protection} -- a reduction in the probability of a loss, as suggested by \cite{EB72}. While such a distinction may be intuitive in models with simple loss distributions or in frequency-severity models, it is sometimes more appropriate to model loss exposure by random variables or distributions and analyze action on that level. Safety measures often influence loss sizes and probabilities together. How useful the distinction of \cite{EB72} is, depends on the modeling framework chosen and the particular application. The term \emph{self-protection} in this paper refers to any activity to reduce physical risk -- including both the size of losses and their probabilities.} levels.
    \item \textbf{Adverse selection}: Agents with larger risks have a higher demand for insurance than safer ones. The degree of the policyholders' risk tolerances cannot be observed by the insurer. The self protection levels of policyholders is not precisely known by the insurer when the contract details are computed.
\end{itemize}

In most papers, cyber insurance is not associated with additional incentives to enhance self protection. In contrast, agents may prefer to buy insurance instead of investments in self protection, i.e., from a welfare perspective, they \textit{underinvest} in security. These observations may be interpreted as an indication that \textit{regulatory interventions} are necessary, such as fines and rebates, mandatory cyber insurance, or minimal investment levels.\footnote{The effect of fines and rebates was studied in papers \cite{Bolot2009}, \cite{liu2014}, \cite{Pal2012}, and \cite{Pal2014}. In the presence of information asymmetries, fines and rebates cannot easily be applied. An alternative regulatory instrument are requirements on minimal investment levels for IT security. However, \cite{shetty2010}, \cite{shettyschwartz2010} and \cite{Schwartz2013} argue against such requirements.}

 \subsubsection{Calibration and Application}\label{sec:critics}
 
 Many questions remain to be answered in future research, since the existing game theoretic models of cyber insurance and cyber security are oversimplified -- thus, not yet fully applicable to real-world data:
 
 \begin{itemize}
     \item \textbf{Simplified network topologies}: In the vast majority of the discussed literature, networks are assumed to be homogeneous. However, agents are typically  heterogeneous  in reality which  substantially alters the cyber ecosystem. Network contagion and cyber loss accumulation are highly sensitive to the topological network arrangement; for example, important determinants are the presence (or the absence) of central hub nodes or clustering effects, see, e.g., \cite{fahrenwaldt2018pricing}. For appropriate risk measurement and management these aspects need to be taken into account explicitly.  
     \item \textbf{Static contagion}: A key feature of cyber risk in networks is the systemic amplification of disturbances. From the insurer's perspective, the contagion dynamics will clearly influence tail risks; an example are catastrophic incidents that affect a large fraction of its portfolio. Such events may be critical in terms of the insurer's solvency. An understanding of cyber losses and an evaluation of countermeasures requires dynamic models of contagion processes.      
    \item \textbf{Constant losses}: In all considered game-theoretic models, the agent's losses are assumed to be constant, i.e., modeled as binary random variables. However, in reality we observe that the severity of instances varies substantially due to the heterogeneity of cyber events, ranging from  mild losses (e.g. malfunctioning of email accounts) to very large losses (e.g. attacks on production facilities, or breakdowns of systems).   
 \end{itemize}
 
Cyber insurance and instruments to control cyber risk depend on the structures of networks, the dynamics of epidemic spread processes, as well as loss models -- and vice versa. These feedback loops need to be properly incorporated in future research. Key ingredients of systemic cyber risks -- the interconnectedness captured by epidemic network models, and strategic interaction described in game-theoretic models -- must be combined.
 
\section{Pricing Cyber Insurance}\label{sec:Pricing}
	
Cyber risk comprises both \emph{non-systemic risk}, further subdivided into \emph{idiosyncratic} and \emph{systematic cyber risk}, cf. Section \ref{sec:ClassicalActuarialApproaches}, and \emph{systemic risk}, cf. Section \ref{sec:systemic}. Classical actuarial pricing, however, relies on the \emph{principle of pooling}, and it is thus applicable for idiosyncratic cyber risks only. For systematic and systemic cyber risk, the appropriate pricing of insurance contracts requires more sophisticated concepts and techniques. A discussion of current industry practice for pricing cyber risks can be found in \cite{romanosky2019}. However, the described approaches  do not yet cover the full complexity of cyber risk such that further (scientific) efforts are necessary. In this section, we explain and suggest suitable pricing techniques\footnote{Another challenge is the insurability of (systemic) cyber risks. Many insurers report that they limit their exposure to this line of business due to a lack of data and models. At the same time, a comparison of rough estimates of supply and potential demand reveals a significant gap in cyber insurance. This is detrimental to agents who are exposed to the risk and do not receive insurance coverage. But insurance companies are also missing out on potentially large business opportunities. In addition to the problems with data and models, however, there is also a fundamental question about the insurability of cyber risk in light of systemic risk. A structured pricing methodology can provide a realistic assessment.} tailored to the three different components of cyber risk. 

\subsection{Pricing of Non-Systemic Cyber Risks}\label{subsec:pricingnonsystemic}
In non-life insurance, contracts are usually signed for one year. At renewal time, the insurer may adjust premium charges as well as terms and conditions, while the policyholder can decide whether or not to continue the contract. Premium calculation thus typically refers to loss distributions on a one-year time horizon. In this section, we adopt this market convention and consider premiums payable annually in advance.\footnote{For simplicity, we assume that interest rates are zero, or alternatively that insurance claims are already discounted.}

In this chapter, we are concerned with a general pricing approach, and we do not restrict ourselves to frequency-severity models. We do, however, adopt some of the notations presented earlier. As introduced in Section \ref{sec:ClassicalActuarialApproaches}, losses and associated premiums are considered in the granularity of cyber risk categories $c\in\{1,\ldots,C\}$ and homogeneous groups $k\in\{1,\ldots, K\}$ of policyholders. Each pair $m=(c,k)$ is called a cyber risk module. In terms of a modular system, the premium per risk category serves as a component for the overall premium. Homogeneous groups -- specified for example in terms of covariates -- correspond to tariff cells, i.e., any policyholder in group $k$ should pay the same premium $\pi^{m,\textrm{non-sys}}$ per risk category $c$. We denote by $n_k$ the number of policyholders in group $k$ and assume that volumes and distributions of risks within a group are identical. Although adopting the previously introduced notation, we do not necessarily consider a frequency-severity approach, but discuss pricing strategies that may also be applied in a more general framework. The methodology is inspired by \cite{WBF2010} and \cite{knispeletal2011}.

To decouple the pricing of idiosyncratic and systematic cyber losses in the absence of systemic risk, one possible approach is to construct a decomposition of the total non-systemic claims amount {$\scal_1^{m,\textrm{non-sys}}$} on a one-year time horizon. This decomposition takes the form
\begin{equation}\label{eq:decomp}
\scal_1^{m,\textrm{non-sys}}=\scal_1^{m, \textrm{idio}}+\scal_1^{m, \textrm{systematic}}
\end{equation}
where the total systematic claims equal $\scal_1^{m, \textrm{systematic}}$ and the term $\scal_1^{m, \textrm{idio}}$ denotes the total idiosyncratic fluctuations around the systematic claims. We explain below how a premium can be computed per risk group. Finally, a smoothing algorithm might be helpful in order to avoid structural breaks between the premiums of risk groups with similar covariates. The terms $\scal_1^{m, \textrm{idio}}$ and $\scal_1^{m, \textrm{systematic}}$ are unique only up to a constant that may be subtracted of one of the terms and added to the other.

In order to obtain a decomposition \eqref{eq:decomp}, we consider the $\sigma$-algebra $\fcal$ that encodes the systematic information. This is, for example, the information that is generated by observing the underlying exogenous stochastic factors. The full information at the time horizon of one year is jointly generated by the $\sigma$-algebra $\fcal$ and idiosyncratic fluctuations, also called technical risks, sometimes explicitly encoded by another $\sigma$-algebra $\mathcal{T}$.  A decomposition  \eqref{eq:decomp} can be obtained by setting
$$\scal_1^{m, \textrm{systematic}} =  \mathbb{E}_\mathbb{P}\left[  \scal_1^{m,\textrm{non-sys}} | \fcal   \right] - \textrm{const}, \quad  \scal_1^{m, \textrm{idio}}  = \scal_1^{m,\textrm{non-sys}} -  \scal_1^{m, \textrm{systematic}} + \textrm{const} ,$$ 
where conditional expectations are computed under the statistical measure $\mathbb{P}$ and where the constant $\textrm{const}$ can freely be chosen. If $\scal_1^{m,\textrm{non-sys}}$ is square-integrable and $\textrm{const}=0$, eq.~\eqref{eq:decomp} corresponds to an orthogonal decomposition in the Hilbert space of square-integrable random variables. An adjustment of the constant might be desirable for allocating the total premium for non-systemic risk to its two components.

\paragraph{Pricing Idiosyncratic Risk} As a special case, we consider the case that idiosyncratic cyber risks in a portfolio of individual claims are conditionally independent given $\fcal$. For homogeneous groups of policyholders, defined in terms of covariates vectors $x^k$, $k\in\{1,\ldots,K\}$, this type of risk is thus subject to pooling of risk, and hence a conditional version of classical actuarial pricing is still applicable. A valuation based on $\fcal$-conditional means with respect to the statistical or real-world measure $\mathbb{P}$ is mathematically justified by a conditional version of the strong law of large numbers. 

More precisely, for each firm $i$ with cyber risk module $m=(c,k)$, annual losses\footnote{Since we are not assuming a frequency-severity model as in Section~\ref{sec:frequencyseverity}, we introduce a slightly different notation to indicate that we are not generally referring to the specific setting discussed in Section~\ref{sec:frequencyseverity}.} {$\hat\scal_1^{m,i}$}, $i\in \bbn$, are identically distributed given $\fcal$, and we suppose that the increments $$\varepsilon^{m,i} := \hat\scal_1^{m,i}- \mathbb{E}_\mathbb{P} [\hat\scal_1^{m,i} |\fil], \quad i\in \bbn,$$ are conditionally independent given $\fil$. Then the average claims amount tends asymptotically\footnote{We recall that $n_k$ denotes the number of policyholders in group $k$.} to the conditionally expected claims amount per policyholder:
$$ \lim_{n_k\uparrow\infty}\tfrac{1}{n_k}\sum_{i=1}^{n_k}\hat\scal_1^{m,i}=\mathbb{E}_\mathbb{P}[\hat\scal_1^{m,1}|\fil]\quad \mathbb{P}\mbox{-a.s.}$$ When setting $\textrm{const}=0$, this is exactly the systematic claims amount for any firm $i$ according to decomposition \eqref{eq:decomp}, suggesting that any premiums per policyholder for idiosyncratic cyber fluctuations should -- for a large number of policyholders $n_k$ in group $k$ -- be equal to zero and only\footnote{This relies on the specific choice $\textrm{const}=0$. When setting $\textrm{const}=0$, idiosyncratic fluctuations will be both positive and negative. From a technical point of view, this does not cause any complications. However, if one needs to require that $\scal_1^{m, \textrm{idio}}\geq 0$, the constant should be suitably adjusted.} the systematic part should be priced. This is analogous to the \emph{net risk premium} in the unconditional setting. But the net risk premium is known to be insufficient! Indeed, in a multi-period model, ruin theory states that ruin of the insurer occurs -- no matter of the initial capital -- in the long run $\mathbb{P}$-a.s. if only the net risk premium is  charged, see, e.g., \cite{mikosch2004non} and the references therein. 

A related result already holds in the one-period setting: Suppose that the premium charged from each firm admits the perfect replication of the systematic part  $\mathbb{E}_\mathbb{P}[\hat\scal_1^{m,1}|\fil]$ (e.g., in a complete financial market). Letting the number of policyholders $n_k$ tend to infinity, the $\fil$-conditional one-period loss probability
$$\mathbb{P}\Bigg( n_k \mathbb{E}_\mathbb{P}[\hat\scal_1^{m,1}|\fil]-\sum_{i=1}^{n_k}\hat\scal_1^{m,i}<0 \,\Big|\, \fil \Bigg)=\mathbb{P}\Bigg(\sum_{i=1}^{n_k}\tfrac{\hat\scal_1^{m,i}-\mathbb{E}_\mathbb{P}[\hat\scal_1^{m,1}|\fil]]}{\sqrt{\operatorname{Var}_\mathbb{P}[\hat\scal_1^{m,1}|\fil]}}>0 \,\Big|\, \fil \Bigg)$$
converges to 50\%, due to the central limit theorem. To stay on the safe side, a safety loading is necessary in addition to the net risk premium. 

In our specific construction, the idiosyncratic part $\scal_1^{m, \textrm{idio}}$  for firm $i$ in \eqref{eq:decomp} equals $\varepsilon^{m,i}$ possessing expectation zero; but in alternative decompositions, a non-zero expectation corresponding to a non-zero net risk premium for the idiosyncratic part would also be admissible. This is an issue of premium allocation between idiosyncratic and systematic cyber risk only, but does not affect the total premium for non-systemic cyber risk, if cash-additive premium principles are used to price idiosyncratic risks.

Classical actuarial premium principles provide explicit safety loadings in a transparent manner, based on the first two moments of the loss distribution:
\begin{itemize}
\item \textbf{Variance principle:} $\pi^{m,\textrm{idio}}=\mathbb{E}[\scal_1^{m, \textrm{idio}}]+a\operatorname{Var}(\scal_1^{m, \textrm{idio}})$ with $a>0$,
\item \textbf{Standard deviation principle:} $\pi^{m,\textrm{idio}}=\mathbb{E}[\scal_1^{m, \textrm{idio}}]+a\sqrt{\operatorname{Var}(\scal_1^{m, \textrm{idio}})}$ with $a>0$.
\end{itemize}
The safety loading parameter $a$ can be chosen for each cyber risk module $m=(c,k)$ separately, for example, depending on the specific loss distribution and the number of contracts $n_k$ in tariff cell $k$. In addition to these simple explicit premium principles, the safety loading can be imposed implicitly, e.g., in terms of convex principles of premium calculation
 including the well-known \emph{exponential principle} or \emph{Wang's premium principle} as special cases, cf. Example \ref{ex:convexpremiumprinciples}.
\paragraph{Pricing Systematic Risk} Systematic cyber incidents affect different firms at the same time -- in contrast to idiosyncratic cyber incidents. Thus, (perfect) pooling of risk is no longer applicable and classical actuarial valuation has to be replaced by a more complex analysis. In the insurance context, systematic risk is not limited to cyber risk only, but also plays a prominent role in the valuation of unit-linked policies or the calculation of the market-consistent embedded value (MCEV) of an insurance portfolio, whereby idiosyncratic actuarial risk and systematic financial market risks must be evaluated together. For an overview on financial pricing methods in insurance we refer to \cite{Bauer2013}.

In this section, we propose a valuation of systematic cyber risk in terms of modern financial mathematics, combining  the principle of \emph{risk-neutral valuation} with the theory of \emph{monetary risk measures}, see \cite{knispeletal2011} for a similar discussion related to the \emph{Market-Consistent Embedded Value} (MCEV) of insurance portfolios. This comprehensive approach requires two different probability measures: While the assessment of risk in terms of a monetary risk measure is based on the real-world measure $\mathbb{P}$ that models the relative frequency with which events occur, valuation of contingent claims in financial mathematics relies on a technical measure $\mathbb{Q}$, called  \emph{risk-neutral measure} or \emph{martingale measure}. In concrete application, tractable models may be obtained by assuming that the systematic one-year losses  $\scal_1^{m, \textrm{systematic}}$ in eq.~\eqref{eq:decomp} is triggered by common risk factors to which all policyholders are jointly exposed. The total claim amount may be viewed as a contingent claim, depending on the evolution of these common factors.
 
Generally speaking, contingent claims are contracts between two or more parties which determine future payments to be exchanged between the parties conditional or contingent on future events. Formally, a contingent claim with payoff at terminal time $t=1$ is described as a random variable. In financial mathematics, the valuation of contingent claims relies on a financial market model on a filtered probability space $(\Omega,\mathcal{F},(\mathcal{F}_t)_{t\in[0,1]},\mathbb{P})$ with a number, say $d+1$, of liquidly traded primary products with price processes $(P^0_t)_{t\in[0,1]}, (P^1_t)_{t\in[0,1]},\ldots, (P^d_t)_{t\in[0,1]}$. The underlying price processes can be modeled either as stochastic processes in discrete time or in continuous time. The asset \lq 0\rq{} plays the role of a num\'{e}raire, i.e., it is used for discounting purposes. A contingent claim $H_1$ maturing at time $t=1$ is called replicable or hedgeable if there exists a self-financing trading strategy\footnote{Intuitively, the self-financing condition means that the portfolio is always rearranged such that on the one hand no additional capital is required and on the other hand no capital is withdrawn.} $\vartheta=(\vartheta_t^0,\vartheta^1_t,\ldots,\vartheta^d_t)_{t\in[0,1]}$ (specifying the number of shares $\vartheta^{i}_t$  of primary products in the portfolio at time $t$) whose terminal wealth $V_1^{\vartheta}$ coincides with the payoff $H_1$ for almost all scenarios. In absence of arbitrage, the price $H_0$ of a replicable contingent claim $H_1$ is unique and equals the \emph{cost of perfect replication}. The calculation of this price can, however, be decoupled from the calculation of the replication strategy itself by the \emph{principle of risk-neutral valuation}. Formally, risk-neutral valuation resembles the classical actuarial valuation in the sense that prices are computed as expectation of future discounted payments. The real-world measure  $\mathbb{P}$ must, however, be replaced by a technical probability measure $\mathbb{Q}$, called \emph{risk-neutral measure} or \emph{martingale measure}. The latter name is motivated by the fact that discounted prices $(P^{i}_t/P^0_t)_{t\in[0,1]}$, $i=0,1,\ldots,d$, must be martingales with respect to $\mathbb{Q}$, i.e., the current discounted price at some time $t$ is the best prognosis of the expected discounted price at a future date $s>t$ given the available information $\mathcal{F}_t$:
$$\mathbb{E}_\mathbb{Q}\left[\tfrac{P^{i}_s}{P^0_s}\bigg\vert \mathcal{F}_t\right]=\tfrac{P^{i}_t}{P^0_t}\quad\mbox{for $0\leq t<s\leq 1$, $i=0,1,\ldots,d$}.$$
 The risk-neutral valuation formula transfers this martingale property to the discounted prices of contingent claims. In particular,
$$H_0=P^0_0\mathbb{E}_\mathbb{Q}\left[\tfrac{H_1}{P^0_1}\right],$$
i.e., the cost of replication can be obtained as expectation of the discounted payoff with respect to any arbitrary (equivalent) martingale measure $\mathbb{Q}$.\footnote{The martingale measure $\mathbb{Q}$ is said to be equivalent to the underlying real-world measure $\mathbb{P}$ if both measures have the same null sets, i.e., for any $A\in\mathcal{F}$ we have $\mathbb{Q}[A]=0$ if and only if $\mathbb{P}[A]=0$. Conceptually, this means that market events that are not relevant with respect to the real-world measure also play no role for the evaluation of contingent claims under $\mathbb{Q}$ and vice versa.}

Markets are, however, typically incomplete\footnote{In absence of arbitrage, incomplete financial market models are characterized by the existence of a whole class of equivalent martingale measures.} in the sense that not every contingent claim can be replicated in terms of liquidly traded primary products. In particular, contingent claims arising from cyber risks cannot be hedged perfectly in the financial market. For non-replicable contingent claims, risk-neutral valuation is still applicable, but now provides -- depending on a whole class of martingale measures -- an interval of prices which are consistent with the absence of arbitrage. Our evaluation of non-replicable contingent claims, however, is based on monetary risk measures and capital requirements. 

Let us denote by $\mathcal{X}$ the set of financial positions with maturity $t=1$ whose risk needs to be assessed.  Mathematically, the family $\mathcal{X}$ is a vector space of real-valued mappings $X_1$ on  $(\Omega, \mathcal{F},\mathbb{P})$ that contains the constants. By sign-convention, negative values of $X_1$ correspond to debt or losses, i.e., the claims amount {$\scal_1^{m, \textrm{systematic}}$} corresponds to the financial position $X_1=-{\scal_1^{m, \textrm{systematic}}}$. A monetary risk measure\footnote{For a rigorous introduction to the theory of risk measures we refer to \cite{FS}.}  $\rho: \mathcal{X} \to \mathbb{R}$ quantifies the risk of a contingent claim that cannot be priced by the cost of perfect replication. Intuitively, a monetary risk measure can be viewed as a capital requirement: $\rho(X_1)$ is the minimal capital that has to be added to the position $X_1$ to make it acceptable, e.g., from the perspective of a financial supervisory authority, a rating agency, or the board of management. To capture the idea that homogeneous risks are assessed in the same way, we assume that $\rho$ is distribution-based, i.e., $\rho(X)=\rho(Y)$ whenever the distributions of $X$ and $Y$ under $\mathbb{P}$ are equal. Prominent examples of distribution-based monetary risk measures are \emph{Value at Risk} ($\var$) and \emph{Average Value at Risk} ($\avar$).\footnote{
For a financial position $X_1$, its Value at Risk at level $\lambda\in (0,1)$ is the smallest monetary amount that needs to be added to $X_1$ such that the probability of a loss does not exceed $\lambda$:
 $$\var_\lambda (X_1) = \inf \{ m\in\mathbb{R}\vert \mathbb{P}[X_1 + m < 0] \leq \lambda\}.$$
 In particular, $\var_\lambda (X_1) = - q^+_{X_1}(\lambda)$,  where $q^+_{X_1}$ is the upper quantile function of $X_1$. The Average Value at Risk, also called Expected Shortfall, at level $\lambda\in (0,1]$ is defined by  $$\operatorname{AVaR}_\lambda (X_1) = \frac{1}{\lambda} \int_0^\lambda \var_\alpha (X_1)\,d\alpha.$$
}

Combining these two approaches, an \emph{algorithm} for the calculation of the premium $\pi^{m,\textrm{systematic}}$ can be summarized as follows {(see also \cite{knispeletal2011})}:
\begin{enumerate}
\item Consider a decomposition of the financial position  $-\scal_1^{m, \textrm{systematic}}=H^m_1+R^m_1$, where
\begin{itemize}
\item $H^m_1$ is a replicable contingent claim with respect to the underlying market model,
\item and $R^m_1$ denotes the residual term.
\end{itemize}
\item Calculate the premium $\pi^{m,\textrm{systematic}}= - H^m_0 + \rho(R^m_1)$, where 
\begin{itemize}
\item
$H^m_0$ equals the cost of perfect replication of $H^m_1$, and the insurance needs to charge its customer the amount $- H^m_0$ for setting up\footnote{Observe that $H_0^{m}$ is typically negative, thus $-H_0^{m}$ positive.} the offsetting position; 
\item $\rho(R^m_1)$ is the premium for $R^m_1$.
\end{itemize}
\end{enumerate}
The decomposition and the premium derived from it may not be unique. From the insurer's perspective, the goal of the decomposition into the summands $(H^m_1,R^m_1)$ is the minimization of the theoretical premium $\pi^{m,\textrm{systematic}}= - H^m_0 + \rho(R^m_1)$ which provides a lower bound\footnote{This lower bound could also directly be computed via a modified risk measure that is constructed according to \cite{FS2002}, see also Chapter 4.8 in \cite{FS}.} for the actual premium charge. The minimization problem apparently involves a trade-off between the cost of replication and the risk of the residual. In practice, it might be reasonable to impose constraints on the decomposition such as upper bounds for $- H^m_0$ and $\rho(R^m_1)$, respectively. Indeed, since the risk of the hedgeable part $H_1^{m}$ can be completely eliminated for the price $- H_0^{m}$, the specification of a bound $\rho(R^m_1)\leq\rho_{\max}$ would already control the overall risk of the systematic cyber losses $\scal_1^{m, \textrm{systematic}}$, in accordance with the company's risk strategy. Conversely, if the insurer's risk budget has not yet been exhausted, it might be helpful to limit the hedging expenses by a bound $-H_0^m\leq \bar H_{\max}$ and to accept the remaining risk $\rho(R^m_1)$.\footnote{{To ensure the existence of a decomposition under constraints, the bounds on risk $\rho_{\max}$ and hedging expenses $\bar H_{\max}$ must satisfy the lower bounds $\rho_{\max}\geq\inf\{\rho(R^m_1)\vert \exists\, (H^m_1,R^m_1)\text{ such that }-\scal_1^{m, \textrm{systematic}}=H^m_1+R^m_1\}$ and $\bar H_{\max}\geq\inf\{-H^m_0\vert \exists\, (H^m_1,R^m_1)\text{ such that }-\scal_1^{m, \textrm{systematic}}=H^m_1+R^m_1\}$, respectively. }}

This concept can be applied for each cyber risk module standalone, but provides additional benefits at portfolio level. If the underlying risk measure $\rho$ is even subadditive (and thus provides incentives for diversification), then the lower bound for the actual premium charge can be further reduced. More precisely, for any given decomposition $-{\mathbb{E}_\mathbb{P}[\hat\scal_1^{m,i}|\fil]}=H^{m,i}_1+R^{m,i}_1$  of the systematic term in eq.~\eqref{eq:decomp} per cyber risk module $m=(c,k)$ and policyholder $i=1,\ldots,n_k$ in group $k$ into a hedgeable part $H^{m,i}_1$ and a residual summand $R^{m,i}_1$, the risk of the residual term of the aggregated systematic risk satisfies
$$\rho\left(\sum_{c=1}^C\sum_{k=1}^K\sum_{i=1}^{n_k} R_1^{m,i}\right)\leq \sum_{c=1}^C\sum_{k=1}^K\sum_{i=1}^{n_k}\rho(R_1^{m,i}),$$ while the costs of perfect replication are additive. Thus, the total premium required at portfolio level is in fact lower than the aggregated premiums:
$$- \sum_{c=1}^C\sum_{k=1}^K\sum_{i=1}^{n_k}H_0^{m,i}+\rho\left(\sum_{c=1}^C\sum_{k=1}^K\sum_{i=1}^{n_k} R_1^{m,i}\right)\leq\sum_{c=1}^C\sum_{k=1}^K\sum_{i=1}^{n_k}(-H_0^{m,i}+\rho(R_1^{m,i})).$$
The diversification effect can be allocated -- according to the insurer's business  strategy -- to the cyber risk modules, yielding a reduction of the lower bound for the actual premium charge per module. 
 
\subsection{Pricing of Systemic Cyber Risks}

Systemic risk is an important issue in cyber insurance. If entities are interconnected, risks may spread and amplify in cyber networks. In addition, this process depends on investments in cyber security and self protection of the agents in the network. Insurance premiums may, in turn, influence investment decisions and thereby modify the safety of the system, cf. Section \ref{sec:systemic}. How to deal with complex cyber systems and the computation of systemic cyber insurance premiums is a topic of current research. 

In this section, we develop some new ideas and introduce a preliminary, stylized approach that builds a bridge between the pricing of cyber insurance contracts and systemic risk measures. We consider $N$ interconnected insurance customers in a cyber network that are also subject to idiosyncratic and systematic risk. For simplicity, we suppose that there exists only a single insurance company that offers $J$ types of contracts. There are two dates, $t=0$ and $t=1$. The initial prices of the insurance contracts, represented by a matrix $\pi= (\pi_{i,j})_{i,j} \in \bbr^{N\times J}$, are $\pi_{i,j}$ where $i=1,2, \dots, N$ denotes the insurance customer and $j=1,2, \dots, J$ the contract type. Each customer $i$ chooses a contract type $j_i$ from this menu and is charged a premium $\pi_{i,j_i}$. Customers decide simultaneously about insurance contracts and their investments into cyber security resulting in random losses $Y^\Pi = (Y^\Pi_i)_{i= 1,2, \dots, N}$ at date $t=1$, the end of the considered period. 

In this setting, we discuss the notion of \emph{systemic premium principles}. Suppose that -- excluding the considered cyber business -- the random net asset value of the insurance firm at date $t=1$ is given by $\tilde E$. Including the cyber contracts, the net asset value\footnote{The interest rate over the considered period is set to $0$ in this example.} of the insurance firm is 
\begin{equation}\label{eg:NEV}
E^\Pi \; = \; \tilde E \;+ \; \sum_{i=1}^N  \pi_{i,j_i} \; - \;  \sum_{i=1}^N Y^\Pi_i.
\end{equation}
The computation of the net asset value implicitly considers network effects that influence losses and the underlying investment decisions of the insurance customers, i.e., the systemic risk inherent in the network.

Systemic premium principles\footnote{The suggested concept of systemic premium principles parallels the notion of systemic risk measures of \cite{feinsteinetal2017}, see also \cite{biaginietal2019}.} refer to the family of premium matrices $\Pi$ that are consistent with solvency requirements or risk limits and admissibility requirements of the insurance company. These can, for example, be formalized in terms of two acceptance sets\footnote{See \cite{FW15} and \cite{FS} for reviews on monetary risk measures.} $\acal^E$  and $\acal^Y$ of monetary risk measures. The solvency condition or risk limit is satisfied, if $E^\Pi \in \acal^E$. An admissibility requirement is that the stand-alone business is viable, i.e.,
\begin{equation}\label{eg:viable}
\sum_{i=1}^N  \pi_{i,j_i} \; - \;  \sum_{i=1}^N Y^\Pi_i  \; \in \; \acal^Y.
\end{equation}
Conditions \eqref{eg:NEV} and \eqref{eg:viable} characterize the systemic premiums, i.e., the family $\mathbb{M}^\Pi$ of admissible premium matrices $\Pi$. 

Idiosyncratic risk and systematic risk can also be priced within this framework. Idiosyncratic risk can be priced by classical actuarial premium principles. This was discussed in Section~\ref{subsec:pricingnonsystemic}. Many premium principles correspond to monetary risk measures that can be encoded by acceptance sets, leading to a framework that is consistent with our suggested approach for pricing systemic risk. The same applies to the residual part of systematic risks that is not replicated. If the insurance firm has access to a financial market that is itself not exposed to systemic risk, it may use this market to partially hedge its exposure. In the absence of systemic risk, this was outlined in Section~\ref{subsec:pricingnonsystemic}. In the current setting, the impact on insurance pricing of trading in financial markets can be included by adjusting the acceptance sets $\acal^E$  and $\acal^Y$ according to \cite{FS2002}, see also Chapter 4.8 in \cite{FS}.

\begin{example}
\label{ex:convexpremiumprinciples}
Solvency regulation varies in different regions of the world. Solvency II in the European Union and the Swiss Solvency Test in Switzerland are based\footnote{To be more precise, the implementation of the regulatory rules are based on Mean-$\var$ and Mean-$\avar$. Details are, e.g., discussed in \cite{weber2018}, \cite{hammetal2020}.} on the risk measures $\var$ and $\avar$, respectively. These risk measures would define the acceptance set $\acal^E$ in our setting.

The acceptance set $\acal^Y$, in contrast, corresponds to a classical premium principle. Indeed, important actuarial premium principles are based on convex risk measures $\rho$ (defined w.\,r.\,t. financial positions) and their acceptance sets $\mathcal{A}$, respectively,\footnote{A monetary risk measure can be recovered from its acceptance set $\mathcal{A}$ via $\rho(X)=\inf\{m\in\mathbb{R}\vert X+m\in\mathcal{A}\}$, i.e., $\rho(X)$ is the smallest capital amount that has to be added such that the financial position becomes acceptable, see, e.g., \cite{FS}.} by choosing 
$$\rho (-L)=\inf\{\pi\in\mathbb{R}\vert \pi-L\in\mathcal{A}\}$$ as a premium for a loss position ${L}\in\mathcal{X} \subseteq L^0_+(\Omega ,\mathcal{F})$.\footnote{For details, see Section 8 in \cite{foellmerknispel2013} and the references therein.}  {Examples\footnote{{We include these examples only for the purpose of illustrating the tractability of the suggested approach that may build on well-known premium principles. Of course, the acceptance set of any other monetary risk measure can also be used. A decision should be made on the basis of what properties are desired.}} of risk measures $\rho$ corresponding to well-known actuarial premium principles are:}
\begin{itemize}
\item \textbf{The family of entropic risk measures:} 
$$\rho_\gamma (X) := \sup_{\mathbb{Q}\in\mathcal{M}_1}\{\mathbb{E}_\mathbb{Q}[-X] - \frac{1}{\gamma} H(\mathbb{Q}\vert \mathbb{P})\}, \quad \gamma\in (0,\infty).$$
Here, $\mathcal{M}_1$ is the set of all probability measures on $(\Omega , \mathcal{F})$, and 
\begin{equation*}
H(\mathbb{Q}\vert \mathbb{P}) := \begin{cases} \mathbb{E}_\mathbb{Q}[\log \frac{d\mathbb{Q}}{d\mathbb{P}}],\quad &\text{if } \mathbb{Q} \ll \mathbb{P}, \\
\infty, \quad &\text{else},\end{cases}
\end{equation*}
is the relative entropy of $\mathbb{Q}$ with respect to a reference measure $\mathbb{P}$, for example, the real-world measure. Using a variational principle for the relative entropy, the entropic risk measure $\rho_\gamma$ takes the explicit form 
$\rho_\gamma(X)=\tfrac{1}{\gamma}\log\mathbb{E}_\mathbb{P}[\exp(-\gamma X)] $
and thus corresponds to the exponential premium principle for the claims amount $Y=-X$. Note that $\rho_\gamma(X)$ is increasing in $\gamma$ and satisfies 
$$\lim_{\gamma\downarrow 0} \rho_\gamma(X)=\mathbb{E}_\mathbb{P}[-X]\quad\mbox{and}\quad
\lim_{\gamma\uparrow \infty} \rho_\gamma(X)=\operatorname{ess\,sup}(-X),$$
i.e., the limiting cases are the negative expected value $\rho(X) = \mathbb{E}_\mathbb{P}[-X]$ (net risk premium) as a lower bound and the maximum loss as an upper bound for premium charges. 

\item \textbf{Distortion risk measures:} For any increasing function $\psi :[0,1]\to [0,1]$ with $\psi (0) = 0$ and $\psi (1) = 1$ the map $c^\psi (A) :=\psi (\mathbb{P}(A))$, $A\in\mathcal{F}$, is called a distortion of a probability measure $\mathbb{P}$. The Choquet integral
$$\rho^\psi (X) := \int (-X) dc^\psi =\int_0^\infty c^\psi[-X>x]\,dx+\int_{-\infty}^0(c^\psi[-X>x]-1)\,dx$$
defines a distortion risk measure, a comonotonic risk measure.  If the distortion function is concave, the distortion risk measure corresponds to Wang's premium principle 
$$\rho^\psi (X) =\int_0^\infty \psi(\mathbb{P}(-X>x))\,dx>\int_0^\infty \mathbb{P}(-X>x)\,dx=\mathbb{E}_\mathbb{P}[-X]$$
that guarantees a non-negative loading for any loss position $Y=-X\geq 0$. In particular, the limiting case $\psi=\operatorname{id}$ again corresponds to the negative expected value which provides a lower bound for the actuarial premium.
\end{itemize}
\end{example}
If we introduce a weak partial order $\leq$ on the space of real-valued $(N\times J)$-matrices $\mathbb{R}^{N\times J}$ by component-wise $\leq$-comparison, the smallest admissible premiums $\bar{\mathbb{M}}^\Pi$ in the family $\mathbb{M}^\Pi$ of admissible premium matrices may be characterized. Although we are dealing only with one insurance firm in our specific construction, the heuristic argument of competitiveness might motivate to focus on premiums in $\bar{\mathbb{M}}^\Pi$ only. Typically, the admissible premium allocations will not be unique. 

A remaining question is the choice of a specific premium allocation. Further criteria or objectives need to be specified for this purpose. We briefly discuss three options:
\begin{itemize}
\item \textbf{Competition:} The heuristic argument of competitiveness might also be used to argue that total premium payments should be as small as possible. This would lead to those allocations where $\sum_{i=1}^N  \pi_{i,j_i} $ is minimal.
\item \textbf{Competitive segments:} If some insurance customers are more price-sensitive and more important than other, one might introduce positive weights $v_i$, $i=1,2,\dots, N$, and focus on allocations with minimal $\sum_{i=1}^N  v_i \pi_{i,j_i} $.
\item \textbf{Performance optimization:} If insurance customers were willing to accept any premium allocation in $\bar{\mathbb{M}}^\Pi$, one could formulate an objective function of the insurance company that determines specific premium allocations. This could be a utility functional or a performance ratio such as RoRaC.
\end{itemize}
A detailed analysis of systemic premium principles in specific models and their statistical and algorithmic implementation are challenging and important questions for future research.

	
	\section{Conclusion and Future Research}\label{sec:Conclusion}
	
In this paper, we provided a comprehensive overview of the current literature on modeling and pricing cyber insurance. For this purpose, we introduced a basic distinction between three different types of cyber risks: \textit{idiosyncratic}, \textit{systematic} and \textit{systemic} cyber risks. Models for both \emph{non-systemic} risk types were discussed within the classical actuarial framework of collective risk models. The (separate) discussion of modeling \emph{systemic} cyber risks then focused on risk contagion among network users in interconnected environments as well as on their strategic interaction effects. Finally, we presented concepts for an appropriate pricing of cyber insurance contracts that crucially relies on the specific characteristics of each of the three risk types.

For both practitioners and academic researchers, modeling and pricing cyber insurance constitutes a relatively new topic. Due to its relevance, the area of research is growing rapidly, but modeling and pricing approaches are still in their infancy. In the introduction, we highlighted four important challenges: \emph{data}, \emph{non-stationarity}, \emph{aggregate cyber risks}, and \emph{different types of risk}.  

Classical actuarial approaches rely heavily on claims \emph{data}.  To date, for cyber insurance these data are sparse and often inaccessible in the actuarial context due to confidentiality issues. As more data becomes available, different modeling approaches  could be more easily tested and evaluated. Therefore, the development of (freely accessible) data collections for cyber risks is an important topic for future research. This requires collaboration between researchers, insurance companies, IT firms, and regulators.  However, due to the evolutionary nature of cyber risks and their \emph{non-stationarity}, the evaluation of data needs to be adaptive, and the relevance of historical information will most likely decrease over time. For this reason, it is important to combine expert opinions supported by advanced modeling techniques with the statistical evaluation of data. 

Our systematic differentiation of risk types - idiosyncratic, systematic and systemic - structures the development of models and the evaluation of data. This facilitates an appropriate consideration of \emph{different types of risks}. We advocate a pluralism of models that provides multiple perspectives in an evolving environment where issues of data availability and data quality remain unresolved. \emph{Aggregate cyber risks} represent an important challenge that needs to be addressed at the systematic and systemic level. In this regard, we have identified the following promising avenues for future research:
\begin{itemize}
     \item \textit{Data on contagion.}  Epidemic network and contagion models require a special kind of data, namely connectivity data for designing realistic network topologies and epidemic spread data for determining epidemic parameter values.The non-stationarity of the cyber environment remains a challenge that must be addressed in this area as well.
    \item \textit{Networks and contagion processes.} The analysis of large-scale network models and epidemic processes is a difficult task.  Developing and improving models and assessment methods is an important research task. Monte Carlo simulations are computationally intensive, and moment closures in mean-field approaches lack estimates of the resulting approximation error. Implementation procedures need to be refined and validated. In addition, realistic loss models are needed for assessing contagious cyber risks.  
    \item \textit{Top-down approaches.} To capture contagion effects in digital networks without rendering the models impossible to handle, a number of top-down approaches has already been developed. These employ population-based models that omit the detailed structure of the underlying networks and processes. However, network topology, e.g., centrality or cluster effects, has a significant role to play. Existing models should therefore be improved via more elaborate refinements that bridge the gap between bottom-up network modeling and population-based top-down approaches.
    \item \textit{Dynamic strategic interaction.} The analysis of strategic interaction effects in cyber models has focused almost exclusively on static frameworks. Such an oversimplification may be inappropriate in environments where systemic spread processes are present. Studying the strategic interaction of network participants in dynamic environments could improve our understanding of the effects of cyber insurance contracts on policyholder behavior and vice versa.
    \item \textit{Multi-layer networks.} Both manufacturing industries and financial operations are now highly dependent on digital technology. Cyberattacks on critical infrastructure pose a systemic threat to modern societies. Such hierarchical systems are characterized by a high degree of interdependence. To understand cyber risks in these structures, analyzing multilayered networks offers a promising approach.
   \item \textit{Pricing systemic cyber risks.} In Section \ref{subsec:pricingnonsystemic}, we outline an approach for pricing systemic cyber risks. It integrates classical valuation concepts and systemic risk measures as a basis for systemic premium principles. Future research must extend the theoretical methodology and apply systemic premium principles in specific models. In addition, statistical and algorithmic techniques need to be developed.
\end{itemize}
 
The list of research tasks that we provide here is not exhaustive. Many further challenges exist. Addressing them will contribute to a more resilient and safer cyber landscape in the future. The evolutionary nature of cyber risk, however, precludes all challenges from ever being finally resolved.
\clearpage 

\begin{appendices}

\appendix
    
    \section{Classification of Cyber Risks}\label{sec:riskcat}

	In this section, we present two exemplary classification approaches of cyber risks from an actuarial perspective: \cite{cro2016} and \cite{zeller2020comprehensive}. For general cyber classification approaches without a specific focus on insurance, we refer to the information security literature, see, e.g., \cite{harrygallagher} and the references therein.
	
\paragraph{\cite{cro2016}}
suggests a classification by manifold factors summarized in Table \ref{tab:CROclassification}. However, due to its granularity, it does not seem very suitable for modeling purposes. Indeed, the classification rather intents to provide a \lq\lq starting point for discussion\rq\rq{} (\cite{cro2016}, p.~24) on a unifying framework for data-gathering purposes. 
  \begin{table}[h]
        \centering
        	 \caption{\cite{cro2016} Classification Overview}
        \vspace{0.2cm}
        \resizebox{\textwidth}{!}{%
	\begin{tabular}{l|l|l|l|l}
		\textbf{Cyber Incident} & \textbf{Event Type} & \textbf{Root Causes}  & \textbf{Actor} & \textbf{Impact Type}\\
		1 System malfunctions/misuse & Operational Risk Categories& A People & 1 Nation states & Business interruption\\
		2 Data confidentiality &  & B External causes& 2 Organized criminals & Data and software loss \\
		3 Data integrity/availability & & C Processes & 3 Hackers & Theft or fraud \\
		4 Malicious activity & & D System & 4 Hacktivists & Cyber ransom and extortion\\
		&&& 5 Insiders & Breach of privacy\\
		&&&& Reputational damage\\
		&&&& Regulatory \& legal defense costs\\
		&&&& Fines and penalties\\
		&&&& Physical asset damage\\
		&&&& and many more, in total: 22 categories
	\end{tabular}}
        \label{tab:CROclassification}
    \end{table}

\paragraph{\cite{zeller2020comprehensive}}
provides a more model-oriented classification of cyber incidents,
see Table \ref{tab:ZellerSchererclassification}. The paper distinguishes between idiosyncratic and systemic incidents.  However, the latter category should, in our view, be further divided into \textit{systematic} and  \textit{systemic} incidents, see the discussion in Section \ref{sec:ClassicalActuarialApproaches}.
	\begin{table}[h]
	    \centering
	    \caption{\cite{zeller2020comprehensive} Classification Examples (see Table 2 therein)}
	    \renewcommand{\arraystretch}{1.5}
	     \vspace{0.2cm}
	    {\footnotesize
	    \begin{tabularx}{\textwidth}{X|XX|XX}
	     & \multicolumn{2}{c}{\itshape Idiosyncratic incidents} & \multicolumn{2}{c}{\itshape Systemic events} \\
	     & Targeted attack & Individual failure & \multicolumn{1}{l}{Untargeted attack} & Mass failure\\\hline
	     {Data Breach (DB)} & Targeted data theft & Individual \mbox{unintended} data disclosure & Data theft through widespread malware / phishing & Unintended data disclosure at cloud service provider\\
	     Business Interruption (BI) & Targeted (D)DoS / Ransomware attack & Disruption of IT system or process through accidental malfunction & Widespread ransomware attack & Cloud service outage disrupting business services \\
	     Fraud / General (FR) & CEO fraud through targeted \mbox{(spear-)}phishing attack & Accidental compromise of database by employee & Widespread ransomware attack or social engineering fraud & Accidental compromise of data stored at cloud service provider
	\end{tabularx}}
	    \label{tab:ZellerSchererclassification}
	\end{table}

\section{Random Network Models}\label{app:randomnet}
Two standard classes of undirected random networks are Erdős–Rényi networks and scale-free networks:
{\begin{itemize}
    \item \textbf{Erdős–Rényi networks.} The simplest random network model was introduced by \cite{Erdos1959}: The Erdős–Rényi network $G_p (N)$ is constructed from a set of $N$ nodes in which each of the possible $N(N-1)/2$ edges is  present independently with equal probability $p$. The resulting degree distribution, i.e., the distribution of the number of neighbors of any node in the network, is binomial, since the probability to create a node of degree $k$ (i.e., with $k$ neighbors) $\mathbb{P}(k)$ is equal to the probability that this node is connected to exactly $k$ other nodes and not connected to the remaining $N-1-k$ nodes of the network:
\begin{equation*}
\mathbb{P}(k) = \binom{N-1}{k} p^k (1-p)^{N-1-k}.
\end{equation*}
For large $N$ and in the limit of constant average degree $(N-1)p\approx Np=:c$,  the binomial distribution can be approximated by a Poisson distribution
\begin{equation*}
\mathbb{P}(k) = e^{-c} \frac{c^k}{k!}.
\end{equation*}
\item \textbf{Scale-free networks.}  Empirical analysis in various research areas suggests that real-world networks exhibit much more heterogeneous degrees than Poisson distributions would suggest. Often a hierarchy of nodes is observable -- with a few nodes of high degree (called \textit{hubs}), and a vast majority of less connected nodes having a relatively low degree. Typically, the degree distribution is approximately \textit{scale-free}, i.e., we have 
\begin{equation*}
    \mathbb{P}(k) \approx a k^{-\lambda}, \qquad a>0,\quad \lambda > 0.
\end{equation*}
A special case with $\lambda = 3$ is given by the \textit{Barabási-Albert model} where a growing network is generated following a preferential attachment rule, see \cite{Barabasi1999} for details.
\end{itemize}}

\section{{Gillespie Algorithm}}\label{app:Gillespie}

{\begin{algorithm}[Gillespie]\label{alg:Gillespie}
\emph{Input:} Initial state of the system $x_0\in E^N$ and initial time $t_0\ge 0$.
	\begin{enumerate}
		\item\emph{(Initialization)} Set the current state $x\to x_0$ and current time $t\to t_0$.
		\item\emph{(Rate Calculation)} For the current state of the system $x$, calculate the sum of rates for all possible transitions $q_x=\sum\nolimits_{i=1}^N q_{x_i}$, where $q_{x_i}$ denotes the rate for a state change of node $i$ according to \eqref{eq:SISSIRrates}.
		\item\emph{(Generate Next Event Time)} Sample the next event time $t_{\text{new}}$ from an exponential distribution with parameter $q_x$.
		\item\emph{(Choose Next Event)} Sample the node $i_{\text{new}}$ at which the next transition occurs: Each node $i=1,\ldots,N$ is chosen with probability ${q_{x_i}}/{q_x}$. \\
		Change the state $x_{i_{\text{new}}}\to y_{i_{\text{new}}}$ according to $\eqref{eq:SISSIRrates}$.
		\item Set $t\to t+t_{\text{new}}$, $x\to (x_1,\ldots,x_{i_{\text{new}}-1},y_{i_{\text{new}}},x_{i_{\text{new}}+1},\ldots,x_N)$ and return to Step 2 until a prespecified stopping criterion is met.
	\end{enumerate}
\end{algorithm}}
	
\section{Moment Closures}\label{app:momentclosures}
This section provides details on moment closures as a measure to solve the Markovian master equation problems \eqref{eq:SIS} and \eqref{eq:SIR}.  

For node $i$, we let $B_i$ be a representative of the Bernoulli random variables $I_i$, $S_i$, or $R_i$ at a certain time $t$. The product $B_{j_1}\new{\cdot\ldots\cdot} B_{j_{k+1}}$, with pairwise different and ordered indices $j_1 < \ldots < j_{k+1} \leq N$, is denoted by $B_J$, $J = \{j_1,\ldots , j_{k+1}\}$. For example, $B_J$ with $J=\{ j_1,j_2,j_3\}$ may denote a triple $I_{j_1}S_{j_2}I_{j_3}$, or $S_{j_1}S_{j_2}I_{j_3}$, etc. 

A moment closure now approximates the moment $\mathbb{E}[B_J]$ by
\begin{equation*}
  \mathbb{E}[B_J] \approx  H(\mathbb{E}[B_{J_1}],\ldots , \mathbb{E}[B_{J_m}]),\quad J_1,\ldots J_m\subset J,\quad\vert J_1\vert ,\ldots , \vert J_m\vert \leq k.
\end{equation*}
Assuming that the single variables $B_i$ are independent leads to the simplest possible moment closure, the \textit{first order independent approximation}, also known as NIMFA in the epidemic literature\footnote{NIMFA is short for ``N-intertwined mean-field approximation'', see \cite{Mieghem2009} for details.}. It is given by
\begin{equation*}
    \mathbb{E}[B_iB_j] = \mathbb{E}[B_i]\mathbb{E}[B_j] + \text{Cov}(B_i,B_j)\approx \mathbb{E}[B_i]\mathbb{E}[B_j].
\end{equation*}
Under this assumption, the full SIS and SIR dynamics are given by the ODE systems of equations \eqref{eq:SIS} and \eqref{eq:SIR}, respectively, when replacing second-order moments with the corresponding product of means. The resulting systems can easily be analyzed by standard techniques from ODE theory.\footnote{For the SIS model, the linear stability condition
$    R_0 = \frac{\tau}{\gamma} < \frac{1}{\hat{\mu}} $
for the infection-free state of the network can be obtained, where $\hat{\mu}$ denotes the spectral radius, i.e., the largest absolute eigenvalue, of the adjacency matrix $A$.} 

However, in certain network structures, the first order independent approximation may yield a large approximation error, see, e.g., \cite{fahrenwaldt2018pricing}.
Hence, more complex approaches to moment closures have been derived. Examples include:
\begin{enumerate}
    \item \textbf{Split closures:} These closures are considered by \cite{fahrenwaldt2018pricing}. The main idea of split closures consists in splitting a set of $k+1$ nodes into two disjoint and non-empty subsets of order $\leq k$:
\begin{equation*}
  H(\mathbb{E}[B_{J_1}],\mathbb{E}[B_{J_2}]) = F(\mathbb{E}[ B_{J_1}])\cdot F(\mathbb{E}[B_{J_2}]), \quad J_1\cap J_2 = \emptyset,\quad J_1\cup J_2 = J, \quad \vert J_1\vert, \vert J_2\vert \leq k,
\end{equation*} 
    with a \emph{mean-field function} $F:[0,1]\to [0,1]$. Different mean-field functions lead to different approximations, e.g.:
\begin{itemize}
    \item \textbf{Independent approximation:} Using the identity as mean-field function, $F(y) = y$, the factorization of the moment of order $k+1$ is done as if the split components were independent:
    \begin{equation*}
         \mathbb{E}[B_J] \approx \mathbb{E}[B_{J_1}] \mathbb{E}[B_{J_2}].
    \end{equation*}
   For the special case $k=1$, this equals the first order independent approximation derived above.
   
   In the SIS model, since 
    \begin{equation*}
     \mathbb{E}[I_J] = \mathbb{E}[I_{J_1}] \mathbb{E}[B_{I_2}] + \text{Cov}(I_{J_1}, I_{J_2})
    \end{equation*}
    and $\text{Cov}(I_{I_1}, I_{I_2})\geq 0$, cf. \cite{Cator2014}, the 
    independent approximation leads to an \textit{upper bound} of infection probabilities.
    
    \item \textbf{Hilbert approximation:} The space of square-integrable random variables forms a Hilbert space with scalar product $\langle Y, Z\rangle := \mathbb{E}[Y\cdot Z]$ and corresponding norm $\Vert Y\Vert := \sqrt{\langle Y,Y\rangle}=\sqrt{\mathbb{E}[Y^2]}$. For $Y,Z\in L^2$, the scalar product defines an angle $\phi$ between the elements:
    \begin{equation}\label{eq:Hilbert}
        \langle Y,Z\rangle = \Vert Y \Vert\cdot \Vert Z\Vert \cdot \cos{\phi}.
    \end{equation}
    Hence, taking the mean-field function $F(y) = \sqrt{y}$, and using the idempotence of Bernoulli random variables, a moment of order $k+1$ can be split into:
    \[\mathbb{E}[B_J] \approx \sqrt{\mathbb{E}[B_{J_1}]} \sqrt{\mathbb{E}[B_{J_2}]}.\]
    Due to \eqref{eq:Hilbert}, the resulting approximation error is low, if the angle $\phi$ is close to $0^\circ$ or $180^\circ$. This observation may be used to determine an optimal split $(J_1,J_2)$ of a given index set $J$. 
    
    In the SIS model, the Cauchy-Schwarz inequality implies that the first order Hilbert approximation leads to a \textit{lower bound} of infection probabilities.
\end{itemize}

To apply these approximations, an appropriate partition scheme $(J_1,J_2)$ for index sets $J$ of order $k+1$ needs to be found. For the SIS model, a first optimal split procedure for both approximation types is suggested in \cite{fahrenwaldt2018pricing}, Algorithm 3.13.

\item \textbf{Kirkwood closures:} These closures constitute the main approach used in the epidemic literature. The underlying idea originates from statistical physics, precisely from the so-called BBGKY hierarchy, which describes the evolution dynamics of an interacting $N$-particle system, originally proposed by \cite{Kirkwood1935}: Considering a set $J\subset\mathcal{V}$ of $k+1$ nodes and the corresponding moment $\mathbb{E}[B_J]$, we only take correlations into account which are stemming from infectious transmissions \textit{over paths of length at most $k-1$}, i.e., passing through a maximum of $k$ nodes. This idea reflects the original statistical physics approach that particle states may be assumed to be independent, if their distance exceeds a certain critical threshold.

Now, assuming the independence of node states which are sufficiently far apart,
the Kirkwood approximation estimates the moment $\mathbb{E}[B_J]$ of Bernoulli random variables with $ J = \{ j_1,\ldots , j_{k+1}\}$ through
\begin{equation*} H(\mathbb{E}[B_{J^1_1}],\ldots , \mathbb{E}[B_{J^1_{m_1}}],\ldots , \mathbb{E}[B_{J^k_{1}}],\ldots , \mathbb{E}[B_{J^k_{m_k}}]) = \prod_{i=1}^k\prod_{\ell=1}^{m_i}\mathbb{E}[B_{J^i_\ell}]^{(-1)^{k-i}} ,
\end{equation*}
where $J^i_\ell\subset J$ denotes a subset of size $i$, $i\le k$, and $\ell\in\{1,\ldots, m_i\}$, i.e., $m_i$ denotes the number of such subsets. A detailed derivation can be found, e.g, in Section V of \cite{Singer2004}. The Kirkwood approximation can be interpreted as generalization of the following scheme:

For $k=1$, states of any two nodes are assumed to be independent, i.e., the approximation equals the first order independent approximation described above.

For $k=2$, we obtain a so-called \textbf{pair-based model}. Here, the system is closed on the level of triplets and \textit{correlations over edges} are considered. In this case, the closure reads 
\begin{equation*}
    \mathbb{E}[B_{j_1}B_{j_2}B_{j_3}] = \frac{ \mathbb{E}[B_{j_1}B_{j_2}] \mathbb{E}[B_{j_1}B_{j_3}]  \mathbb{E}[B_{j_2}B_{j_3}]} { \mathbb{E}[B_{j_1}] \mathbb{E}[B_{j_2}] \mathbb{E}[B_{j_3}]}.
\end{equation*}

Two different cases for the node triplet $\{j_1, j_2, j_3\}$ may be considered: For \textit{closed} triplets, i.e., triplets in which edges exist between all pairs of nodes (triangles), node states are pairwise correlated, and hence, the closure cannot be reduced. In contrast, for an \textit{open} triplet only consisting of edges $(j_1,j_2)$ and $(j_2,j_3)$, the states of nodes $j_1$ and $j_3$ are assumed to be independent, and therefore, the closure may be reduced to
\begin{equation*}
      \mathbb{E}[B_{j_1}B_{j_2}B_{j_3}] = \frac{ \mathbb{E}[B_{j_1}B_{j_2}] \mathbb{E}[B_{j_2}B_{j_3}]} { \mathbb{E}[B_{j_2}]}.
\end{equation*}
This equals the \textit{exact result} for $\mathbb{E}[B_{j_1}B_{j_2}B_{j_3}]$ \textit{under the independence assumption}, using Bayes' theorem.

Thus, in the SIR Markov model, \textit{exact closures} can be derived when considering \textit{cut-vertices} $i$, i.e., nodes connecting two otherwise disconnected subgraphs $G_1$ and $G_2$ of the network: If $i$ is in the susceptible state of the SIR model, the infection has not yet passed through this node. Hence, the infection processes in the subgraphs $G_1$ and $G_2$, that are connected solely through $i$, are independent of each other, see, e.g., \cite{Kiss2015}. This result in particular applies to tree graphs, where, by definition, all nodes with degree higher than one are cut-vertices and all triplets are open with $B_{j_2} = S_{j_2}$. For tree networks, the SIR pair-based model is thus exact.

\end{enumerate}

\section{{Estimation of (Cyber) Epidemic Network Models}}\label{app:NetworkEstimation}

Statistical estimation relies first on an underlying statistical model that specifies a range of probabilistic mechanisms that might have generated the data, and second on the observable data. Both components, the model framework and the available data, define the statistical challenge that needs to be addressed. We briefly review some work that focuses on inference for SIS, SIR, or related models. In all cases, the resulting propagation process is simply denoted by $(X(t))_{t\ge 0}$, although we consider different models. The specification of the interaction between entities in the underlying probabilistic mechanisms in the statistical model can be interpreted as a graph $G$ in this framework. The graph $G$ may simply be encoded by an adjacency matrix in some models; in other, heterogeneous models, it might be described as a weighted graph corresponding to a matrix with entries different from $0$ and $1$ that encodes the interaction in the underlying statistical model.

In the context of statistical inference, some parameters of the interaction dynamics are unknown, such as overall infection and recovery rates, but in some problems the interaction graph $G$ might still be known a priori, while in others the graph must be inferred on the basis of available data. We classify the estimation approaches for (cyber) epidemic network models of SIS-, SIR- or related type roughly in the following way:

First, we distinguish if on the level of the underlying statistical model the interaction graph $G$ is known; second, on the level of the data, we differentiate two situations, i.e., the realization of the infection process $X$ might be directly observable or, alternatively, only some related data might be observable, while the spread process itself is hidden. We refer to Sections \ref{subsubsec:networks} and \ref{subsubsec:spreadprocess} for background on spread models. In this section, we provide a brief review of some papers that belong to the following possible categories:

\begin{itemize}
	\item[{$- -$}] {$G$ unknown \&  $(X(t))_{\ge 0}$ not directly observable:}
	
	{In a cyber epidemic network context, \cite{Antonio2021} propose a graph mining approach in a generalized SIS network model (in which infection rates are heterogeneous and self-infection is possible) where the process $X$ is not directly observable, but only auxiliary communication data is available. A filtering mechanism is applied that deletes low-weighted edges below a minimum weight threshold. However, the model is not yet calibrated with real-world data.  For readers interested in more general (inverse) problems, we refer to the book by \cite{Kolaczyk2009}.}
	
	\item[{$++$}] {$G$ known \& $(X(t))_{t\ge 0}$ observable:}
	
	{If $G$ is known and the realization of $X$ is observable, the statistical problem boils down to inference of the epidemic parameters $\tau$ and $\gamma$ of the epidemic spread model in the case of a Markovian SIR network model. This is discussed in Section 6.1 of \cite{britton2020epidemic}; the estimation can be implemented using a maximum-likelihood approach.}
	
	\item[{$+{(-)}$}]{$G$ known \& only terminal individual information on $(X(t))_{t\ge 0}$ observable:}
	
	{\cite{britton2020epidemic}, Section 6.1., discusses the case that $G$ possesses fully connected subgraphs in the case of a Markovian SIR network model, i.e., a so-called household structure. In this case, a maximum-likelihood approach to estimate the epidemic parameters is still feasible, if only the realization of $X$ at a final date is observable, but not its whole evolution. However, the author emphasizes that, without making this specific structural assumption, epidemic parameter estimation is not straight-forward for arbitrary known graphs, if e.g. only the realization of the final number of infections is observable. One approach to overcome this difficulty could thus be to gather some additional time-dependent spread data.}
	
	\item[{$(+)(+)$}] {$G$ unknown, but network model class fixed \& only individual recovery information on $(X(t))_{t\ge 0}$ observable:}
	
	{Another example are random graph models. For example, \cite{Britton2002} develop a Bayesian approach in a Markovian SIR model to estimate the epidemic parameters $\tau$ and $\gamma$ jointly with the connection probability $p$ in Erdős–Rényi type networks, if the spread process $(X(t))_{t\ge 0}$ or only the individual recovery processes $(R_i(t))_{t\ge 0}=(\mathbbm{1}_{X_i(t)=R})_{t\ge 0}$, $i=1,\ldots,N$, are observable (see also \cite{Groendyke2011} for a generalization to the SEIR model and Gamma-distributed infection arrival times).
	Samples from the posterior distribution can be generated using MCMC methods.}

	\item[{$(-)(-)$}]{$G$ unknown, but set of possible network model classes fixed \& only aggregate infection information on $(X(t))_{t\ge 0}$ observable:}
	
	{Often the \emph{individual} time-dependent spread data is not observable, but only the evolution of the \emph{aggregate} number of infections over time is known. To overcome this issue, for example, in a Markovian SIS framework, \cite{Lauro2020} suggest a so-called birth-death process approximation (see also \cite{Zerenner2022} for an extension of this approach to the question of forecasting an ongoing epidemic). Such birth-death processes keep track of the number of infected nodes at population level and thus present an approximation of the original Markovian spread processes in a reduced dimension. \cite{Lauro2020} provide a Bayesian estimation framework in which the epidemic and network parameters for certain random network classes can jointly be estimated; in particular, the method is able to identify the most likely network class out of a regular, Erdős–Rényi, or Barabási-Albert model.}
	
	\item[{$-+$}] {$G$ unknown \& $(X(t))_{\ge 0}$ observable:}
	
	{In the previously discussed approaches with an observable epidemic process, the network $G$ is (partially) known -- at least it belongs to a set of random network classes. How can one proceed if no prior information is available on the network on the level of the statistical model, but the realization of the infection spread process is observable? One suggestion is a cascade approach in (possibly non-Markovian) SI-models (also: activation/information diffusion models). The goal is to infer the network structure under the assumption that the cascades of infections are fully observable using a likelihood approach. The proposed methods in the literature mostly differ in their assumptions on the spread process. For example, \cite{Myers2010} and \cite{Gomez2012} assume homogeneous parametric infection arrival distributions (see also \cite{Gomez2014} for dynamically evolving networks), while \cite{Du2012} do not impose distributional assumptions, but propose a kernel estimation method to estimate the network structure.}
\end{itemize}

{Our overview is not meant to be exhaustive, but intends to highlight different perspectives possibly implied by the structure of a specific application. We also refer to the surveys \cite{Brugere2018} or \cite{kolaczyk2014statistical}. The current literature on (epidemic) network estimation is fragmented with each approach tackling only a specific problem at a time. A unifying methodology does not yet exist -- and is maybe also not realistic to expect. Many questions remain open, see, e.g., the discussion in \cite{britton2020epidemic}. Statistics for (cyber) epidemic network models will most likely continue to be a very active field of research in the future.}

\end{appendices}

	
	\printbibliography
	
\end{document}